\documentclass[amsmath,amssymb,prc,final,twocolumn]{revtex4-1}
\usepackage{graphics,epsfig}
\usepackage{graphicx}
\usepackage[varg]{txfonts} 
\usepackage[latin1]{inputenc}
\newcommand{\cm}{\mathrm{c\!\:\!.m\!\:\!.}}
\begin{document}
\title{%
Coulomb Effects in Few-Body  Reactions
}%
\author{%
A. Deltuva\thanks{\email{deltuva@cii.fc.ul.pt}} 
}
%
\affiliation{
Centro de F\'{\i}sica Nuclear da Universidade de Lisboa,
P-1649-003 Lisboa, Portugal
}
%
\begin{abstract}
The method of screening and renormalization is used to include
the Coulomb interaction between the charged particles in the momentum-space
description of three- and four-body nuclear reactions.
The necessity for the renormalization of the scattering amplitudes and 
the reliability of the method is demonstrated.
The Coulomb effect on observables is discussed.
\end{abstract}

\maketitle
%
%
%
\section{Introduction \label{sec:intro}}

The inclusion of the long-range Coulomb interaction in the description 
of the three- and four-particle scattering is a challenging task in 
theoretical few-body
nuclear physics. The long range of the Coulomb potential 
prevents the direct application of the standard scattering theory.
There is a number of suggestions how to overcome this difficulty;
most of them are based on the configuration-space
framework \cite{kievsky:96a,chen:01a,ishikawa:03a,doleschall:05a} and are
limited to energies below three-body breakup threshold, while the
others \cite{alt:04a,kadyrov:05a,kadyrov:09a,oryu:06a}  have not matured yet
into practical applications. Up to now only few approaches led
to the results above three-body breakup threshold. 
Those are configuration-space calculations
for proton-deuteron ($p$-$d$) elastic scattering
using the Kohn variational principle \cite{kievsky:01a} and
the screening and renormalization method in the framework of momentum-space
integral equations \cite{alt:94a,alt:02a,deltuva:05a,deltuva:05d};
the latter approach will be discussed here in more details.
Very recently $p$-$d$ results above three-body breakup threshold 
were also obtained using
modified Faddeev equation in configuration space together with the dumping
of particular Coulomb contributions \cite{ishikawa:09a}.

\section{Method of screening and renormalization \label{sec:th} }

In nature the Coulomb potential $w_C$ is always screened at large distances.
The comparison of the data from typical nuclear physics experiments
and  theoretical  predictions with full Coulomb  is meaningfull
only if the full and screened Coulomb become
\emph{physically indistinguishable}. This was proved
 in Refs.~\cite{taylor:74a,semon:75a}
where the screening and renormalization method for the scattering of two 
charged particles was proposed.
We base our treatment of the Coulomb interaction  on that idea.

The standard scattering theory is formally applicable to the screened
Coulomb potential $w_R$, i.e.,  the Lippmann-Schwinger equation yields
the two-particle transition matrix 
\begin{equation} \label{eq:tr}
t_R = w_R + w_R g_0 t_R
\end{equation}
where $g_0$ is the free resolvent.
 In the configuration-space representation we choose the screened
Coulomb potential as
\begin{equation} \label{eq:wr}
w_R(r) = w_C(r)\; e^{-(r/R)^n}
\end{equation}
with $R$ being the screening radius
and $n$ controlling the smoothness of the screening.
It was proven in Ref.~\cite{taylor:74a}  that
in the limit of infinite screening radius $R$ the on-shell screened Coulomb 
transition matrix (screened Coulomb scattering amplitude)
$\langle \mathbf{p}'| t_R |  \mathbf{p} \rangle$ with $p'=p$,
renormalized by the infinitely oscillating phase factor
$z_R^{-1}(p) = e^{2i\phi_R(p)}$, approaches the  full Coulomb amplitude 
 $\langle \mathbf{p}'| t_C |  \mathbf{p} \rangle$
in general \emph{as a distribution}, i.e.,
\begin{equation} \label{eq:taylor}
\lim_{R \to \infty} z_R^{-1}(p)
\int d^2 \mathbf{\hat{p}} \langle \mathbf{p}'| t_R |  \mathbf{p} \rangle
\varphi(\mathbf{\hat{p}}) =
\int d^2 \mathbf{\hat{p}} \langle \mathbf{p}'| t_C |  \mathbf{p} \rangle
\varphi(\mathbf{\hat{p}})
\end{equation}
for any test function $\varphi(\mathbf{\hat{p}})$ with the properties
given in Ref. \cite{taylor:74a}; in particular, it must vanish for forward
scattering, i.e., $\varphi(\mathbf{\hat{p}}') = 0$.
Reference \cite{taylor:74a} uses partial wave expansion,
which for the  full Coulomb amplitude itself converges only as a
distribution, and it therefore is unable to make any conclusions on the
possible pointwise convergence.
However, as argued already in Ref.~\cite{taylor:74a},
the convergence of the renormalized screened Coulomb scattering
 amplitude to the full Coulomb amplitude in the sense of distributions
is sufficient for the description of physical observables. For a fixed
final state observation direction $ \mathbf{\hat{p}}'$ the cross section
is determined not directly by the scattering amplitude, but
by the outgoing wave packet 
\begin{eqnarray}
\varphi_f(\mathbf{p}') & = & \int d^3\mathbf{p} 
\langle \mathbf{p}'| S |  \mathbf{p} \rangle \varphi_i(\mathbf{p}) \\
& \sim &  
\int d^2 \mathbf{\hat{p}} \langle \mathbf{p}'| t_R |  \mathbf{p} \rangle
\varphi_i(\mathbf{p}) |_{p=p'}
\end{eqnarray}
that is related to the initial wave packet $\varphi_i(\mathbf{p})$
by the $S$-matrix or the scattering amplitude.
Thus, in the step from  the scattering amplitude to the
cross section one has to go through the conceptual exercise of
averaging the scattering amplitude over the initial state physical wave
packet being peaked around the experimental beam momentum
$\mathbf{{p}}_i$. Thus,  on the rhs of Eq.~(\ref{eq:taylor})
 $\langle \mathbf{p}'| t_C |  \mathbf{p}_i \rangle$
itself is picked out in that average.
In addition, the outgoing wave packet is never observed in
the forward direction, i.e., $ \mathbf{\hat{p}}' \ne \mathbf{\hat{p}}_i$; 
the necessary
property  $\varphi_{i}(\mathbf{p}') = 0$ can
therefore always be fulfilled by the sharpness of the initial wave packet.
In the practical calculations \cite{alt:02a,deltuva:05a} the above averaging
is carried out implicitly, replacing the renormalized screened Coulomb 
amplitude in the $R \to \infty$ limit by the full one, i.e.,
\begin{equation} \label{eq:taylor2}
\lim_{R \to \infty} z_R^{-1}(p)
 \langle \mathbf{p}'| t_R |  \mathbf{p} \rangle \to
 \langle \mathbf{p}'| t_C |  \mathbf{p} \rangle.
\end{equation}
Since $z_R^{-1}(p)$ is only a phase factor,
the above relations indeed demonstrate that the physical observables
become insensitive to screening provided it takes place at sufficiently
large distances $R$ and, in the $R \to \infty$ limit,
coincide with the corresponding quantities referring to the full Coulomb.
Furthermore,
renormalization by $ z_{R}^{-\frac12}(p_i)$ in the $R \to \infty$ limit relates
also the screened and full Coulomb wave functions \cite{gorshkov:61}, i.e.,
\begin{equation} \label{eq:gorshkov}
\lim_{R \to \infty} (1 + g_0 t_R) |\mathbf{p} \rangle z_R^{-\frac12}(p)
    =  |\psi_C^{(+)}(\mathbf{p}) \rangle.
\end{equation}

\subsection{Three-particle scattering}

The screening and renormalization method based on the above relations
can be extended to more complicated systems, albeit with some limitations.
The systems of two- and three-particles  interacting via pairwise strong
short-range and screened Coulomb potentials, $v_\alpha$ and $w_{\alpha R}$,
$\alpha = 1,2,3$,  is considered in 
Refs.~\cite{deltuva:05a,deltuva:05d,deltuva:08c}.
Here we extend our treatment to a more general case where an irreducible
three-body force 
\begin{equation}
 V_{(3)} = \sum_{\alpha=1}^3 u_\alpha
\end{equation}
is present; it is  decomposed into three terms $u_\alpha$.
The full resolvent
\begin{equation} \label{eq:GR1}
G^{(R)} = [E + i0 - H_0 - \sum_\gamma (v_\gamma + u_\gamma + w_{\gamma R})]^{-1},
\end{equation}
with $H_0$ being the three-particle kinetic energy operator
and $E$ the available energy may be decomposed into channel resolvents
\begin{equation} \label{eq:GRa}
 G^{(R)}_\alpha = (E + i0 - H_0 - v_\alpha - w_{\alpha R})^{-1},
\end{equation}
and the multichannel three-particle transition operator
$U^{(R)}_{\beta \alpha}$ according to
\begin{equation} \label{eq:GR2}
  G^{(R)} = \delta_{\beta \alpha}  G^{(R)}_\alpha  +
  G^{(R)}_\beta   U^{(R)}_{\beta \alpha}   G^{(R)}_\alpha ;
\end{equation}
all operators depend parametrically on the Coulomb screening radius $R$.
The full multichannel transition matrix $U^{(R)}_{\beta \alpha}$ 
for elastic and rearrangement scattering
is calculated from the integral equation \cite{deltuva:09e}
\begin{eqnarray} \nonumber
 U^{(R)}_{\beta \alpha} & = {} & \bar{\delta}_{\beta \alpha} G_0^{-1} 
+ \sum_\gamma  \bar{\delta}_{\beta \gamma} 
T^{(R)}_\gamma G_0 U^{(R)}_{\gamma \alpha} \\ &  & + \; u_\alpha 
+ \sum_\gamma  u_\gamma G_0(1+T^{(R)}_\gamma G_0)  U^{(R)}_{\gamma \alpha}
\label{eq:Uba}
\end{eqnarray}
that is a generalisation of the Alt, Grassberger, and Sandhas (AGS) 
equation \cite{alt:67a} in the presence of the three-body force.
$G_0 = (E+i0 - H_0)^{-1}$ is the free resolvent,
$\bar{\delta}_{\beta \alpha} = 1 - {\delta}_{\beta \alpha}$, and
the two-particle transition matrix is derived from the full
channel interaction $v_\alpha + w_{\alpha R}$, i.e.,
 \begin{equation} \label{eq:TR}
   T^{(R)}_\alpha  =  (v_\alpha + w_{\alpha R}) +
     (v_\alpha + w_{\alpha R})  G_0 T^{(R)}_\alpha.
  \end{equation}
The on-shell matrix elements
$\langle b_{\beta}\mathbf{q}'|U^{(R)}_{\beta \alpha}
|b_{\alpha}\mathbf{q}\rangle$
are  amplitudes (up to a factor) for  elastic 
($\beta = \alpha$) and rearrangement ($\beta \neq \alpha$) scattering.
The channel states $|b_{\alpha}\mathbf{q}\rangle$  are the
eigenstates of the corresponding channel Hamiltonian $H_\alpha = H_0 + v_\alpha$
with the energy eigenvalue $E$. $|b_{\alpha}\mathbf{q}\rangle$ is
a product of the bound state wave function $|b_\alpha \rangle$
for the pair $\alpha$ and the
plane wave with the relative particle-pair  $\alpha$ momentum
$\mathbf{q}$; the dependence on the discrete quantum numbers is suppressed 
in our notation.

In order to isolate the screened Coulomb contributions to the transition 
amplitude that diverge in the infinite $R$ limit, we use a
decomposition of the full resolvent into alternative channel resolvents
\begin{equation}
    G_{\alpha R} =
    (E+i0 - H_0 - v_\alpha - w_{\alpha R} - W^{\cm}_{\alpha R})^{-1},
\end{equation}
where $W^{\cm}_{\alpha R}$ is the screened Coulomb potential
between the spectator particle $\alpha$
and the center of mass (c.m.) of the remaining pair.
The same screening function is used for both Coulomb potentials
$w_{\alpha R}$ and $W^{\cm}_{\alpha R}$.
The corresponding transition matrix
\begin{equation} \label{eq:Tcm}
T^{\cm}_{\alpha R}  = W^{\cm}_{\alpha R} +
W^{\cm}_{\alpha R} G^{(R)}_{\alpha}  T^{\cm}_{\alpha R},
\end{equation}
is a two-body operator and therefore its on-shell and half-shell 
behaviour in the limit $R \to \infty$
is given by Eqs.~(\ref{eq:taylor}) and (\ref{eq:gorshkov}). 
It relates the channel resolvents as
\begin{equation}
    G_{\alpha R} =  G^{(R)}_{\alpha} + 
    G^{(R)}_{\alpha} T^{\cm}_{\alpha R} G^{(R)}_{\alpha}.
\end{equation}
Thus, the full resolvent can alternatively be decomposed into
 \begin{eqnarray} \label{eq:GR4b}
G^{(R)} & = &  \delta_{\beta \alpha}   G_{\alpha R} +
    G_{\beta R} \tilde{U}^{(R)}_{\beta\alpha} G_{\alpha R} \\
   & = {} & \delta_{\beta \alpha} G^{(R)}_{\alpha} +
 G^{(R)}_{\beta} \delta_{\beta \alpha} T^{\cm}_{\alpha R}
 G^{(R)}_{\alpha} \nonumber \\
 & &
 + G^{(R)}_{\beta}  [1 + T^{\cm}_{\beta R} G^{(R)}_{\beta}]
 \tilde{U}^{(R)}_{\beta\alpha} 
[1 + G^{(R)}_{\alpha} T^{\cm}_{\alpha R}]  G^{(R)}_{\alpha}, \quad
  \end{eqnarray}
where the  reduced transition operator $ \tilde{U}^{(R)}_{\beta\alpha}(Z)$
may be calculated through the integral equation
\begin{eqnarray} \nonumber
 \tilde{U}^{(R)}_{\beta\alpha} & = {} &
\bar{\delta}_{\beta \alpha} (G_{\alpha R}^{-1} + v_{\alpha}) + u_\alpha +
{\delta}_{\beta \alpha} \mathcal{W}_{\alpha R}  \\ & &
+ \sum_\gamma (\bar{\delta}_{\beta \gamma} v_\gamma + u_\gamma +
{\delta}_{\beta \gamma} \mathcal{W}_{\beta R})
G_{\gamma R} \tilde{U}^{(R)}_{\gamma\alpha}
\label{eq:tU}
\end{eqnarray}
which is driven by the strong two- and three-body 
potentials $v_\alpha$ and $u_\alpha$ and the  potential
of three-body nature
$\mathcal{W}_{\alpha R} = \sum_{\gamma}
( \bar{\delta}_{\alpha \gamma} w_{\gamma R} -
\delta_{\alpha\gamma} W^{\cm}_{\gamma R} ) $.
This potential $\mathcal{W}_{\alpha R}$
accounts for the difference between the direct Coulomb interaction $w_{\gamma R}$
and the auxiliary one  $W^{\cm}_{\gamma R}$
that takes place between the charged particle and the c.m. of the
remaining pair.
When calculated between on-shell screened Coulomb states,
$\tilde{U}^{(R)}_{\beta\alpha}$ is of
short-range, even in the infinite $R$ limit.
Equation~(\ref{eq:GR4b}), together with Eq.~(\ref{eq:GR2}),
 gives a relation between full and reduced three-particle transition
operators, i.e.,
\begin{eqnarray} 
    U^{(R)}_{\beta \alpha} &=& \delta_{\beta\alpha} T^{\cm}_{\alpha R}
    +  [1 + T^{\cm}_{\beta R} G^{(R)}_{\beta}] 
    \tilde{U}^{(R)}_{\beta\alpha}  
          [1 + G^{(R)}_{\alpha} T^{\cm}_{\alpha R}] \quad
\label{eq:U-T} \\
 &=& \delta_{\beta\alpha} T^{\cm}_{\alpha R}
+ (U^{(R)}_{\beta \alpha} - \delta_{\beta\alpha} T^{\cm}_{\alpha R}).
\label{eq:U-T2}
\end{eqnarray}
Thus, the three-particle transition operator $U^{(R)}_{\beta \alpha}$
has a long-range part $\delta_{\beta\alpha} T^{\cm}_{\alpha R}$ whereas the
remainder  $U^{(R)}_{\beta \alpha} - \delta_{\beta\alpha} T^{\cm}_{\alpha R}$
is a short-range operator that  is externally distorted
due to the screened Coulomb waves generated by
$[1 + G^{(R)}_{\alpha} T^{\cm}_{\alpha R}]$.
On-shell, both parts do not have a proper limit as $R \to \infty$ but
the limit exists after renormalization by an appropriate phase factor,
yielding  the transition amplitude for full Coulomb
\begin{eqnarray} \nonumber
&&    \langle  b_\beta \mathbf{q}' | U^{(C)}_{\beta \alpha}
    |b_\alpha \mathbf{q}\rangle     = 
    \delta_{\beta \alpha}
    \langle b_\alpha \mathbf{q}' |T^{\cm}_{\alpha C}
    |b_\alpha \mathbf{q} \rangle  \\ & &  
 +   \lim_{R \to \infty} [  Z^{-\frac{1}{2}}_{\beta R}(q')
    \langle b_\beta \mathbf{q}'  |
            ( U^{(R)}_{\beta \alpha}  
             - \delta_{\beta\alpha} T^{\cm}_{\alpha R})
            |b_\alpha \mathbf{q} \rangle
             Z^{-\frac{1}{2}}_{\alpha R}(q) ]. \quad
\label{eq:UC2}
\end{eqnarray}
The first term on the right-hand side of Eq.~(\ref{eq:UC2}) is known 
analytically \cite{taylor:74a}; it corresponds to the  particle-pair $\alpha$
full Coulomb transition amplitude 
that results from the implicit renormalization of $T^{\cm}_{\alpha R}$ 
according to Eq.~(\ref{eq:taylor2}).
The $R \to \infty$ limit for the remaining part
$( U^{(R)}_{\beta \alpha} - \delta_{\beta\alpha} T^{\cm}_{\alpha R})$
of the multichannel transition matrix is performed numerically; 
due to the short-range nature of this term, as demonstrated in 
Eq.~(\ref{eq:U-T}), the convergence with the increasing screening radius $R$
is fast and the limit is reached with sufficient accuracy at
finite $R$; furthermore, it can be calculated using the partial-wave expansion.
We emphasize that  Eq.~(\ref{eq:UC2}) is by no means an approximation
since it is based on the obviously exact identity (\ref{eq:U-T2})
where the  $R \to \infty$ limit for each term exists and
is calculated separately.

The renormalization factor for  $R \to \infty $ is a diverging phase factor
  \begin{equation}
    Z_{\alpha R}(q) = e^{-2i \Phi_{\alpha R}(q)},
  \end{equation}
  where $\Phi_{\alpha R}(q)$, though independent of the particle-pair relative
  angular momentum $l$ in the infinite $R$ limit, may be realized by
  \begin{equation}    \label{eq:phiRl}
    \Phi_{\alpha R}(q) = \sigma_l^{\alpha}(q) -\eta_{lR}^{\alpha}(q),
  \end{equation}
with the diverging screened Coulomb phase shift $\eta_{lR}(q)$
corresponding to standard boundary conditions
and the proper Coulomb one $\sigma_l(q)$ referring to the
logarithmically distorted proper Coulomb boundary conditions.
For the screened Coulomb potential of Eq.~(\ref{eq:wr})
the infinite $R$ limit of $\Phi_{\alpha R}(q)$ is known analytically,
\begin{equation}  \label{eq:phiRlln}
  \Phi_{\alpha R}(q) =  \mathcal{K}_{\alpha}(q)[\ln{(2qR)} - C/n],
  \end{equation}
where  $C \approx 0.5772156649$ is the Euler number and
$\mathcal{K}_{\alpha}(q)$ is the Coulomb parameter. 
The form of the renormalization phase $\Phi_{\alpha R}(q)$ to be used in the
actual calculations with finite screening radii $R$ is not unique,
but the converged results show independence of
the chosen form of $\Phi_{\alpha R}(q)$.

For breakup observables we follow a very similar strategy where the starting
point is the AGS-like equation for the breakup operator 
\begin{eqnarray} \nonumber
 U^{(R)}_{0 \alpha} & = {} & G_0^{-1} 
+ \sum_\gamma  T^{(R)}_\gamma G_0 U^{(R)}_{\gamma \alpha} \\ &  & + \; u_\alpha 
+ \sum_\gamma  u_\gamma G_0(1+T^{(R)}_\gamma G_0)  U^{(R)}_{\gamma \alpha}
\label{eq:U0a} 
\end{eqnarray}
and its relation to the full resolvent, i.e.,
\begin{equation} \label{eq:GU0}
G^{(R)} = G_0 U^{(R)}_{0\alpha} G^{(R)}_\alpha .
\end{equation}
In the same spirit, we introduce auxiliary  Coulomb resolvent
  \begin{equation}
    G_R =  (E+i0 - H_0 -  \sum_\gamma  w_{\gamma R})^{-1}.
  \end{equation}
that keeps only the screened Coulomb interaction.
The proper three-body Coulomb wave function and its relation to the
three-body screened Coulomb wave function generated by $G_R$ is, in general, unknown.
This prevents the application of the screening and renormalization method to the
reactions involving three free charged particles (nucleons or nuclei) in the final state.

However, in  the system of two charged particles and a neutral one,
only the channel $\gamma = \rho$, corresponding to the correlated pair
of charged particles, contributes to $G_R$ which simplifies to
\begin{eqnarray}
  G_R = {} & G_0 + G_0 T_{\rho R} G_0, \\
  T_{\rho R} = {} & w_{\rho R} + w_{\rho R} G_0 T_{\rho R},
\end{eqnarray}
making channel $\rho$ the most convenient choice for the
description of the final breakup state.
Thus, for the purpose of breakup, a decomposition of the full
resolvent, alternative to Eq.~(\ref{eq:GU0}) is
\begin{eqnarray}
    G^{(R)} & = {} &
    G_{R} \tilde{U}^{(R)}_{0\alpha} G_{\alpha R} \\
     & = {} &  G_{0}  [1 + T_{\rho R} G_{0}]
    \tilde{U}^{(R)}_{0\alpha} 
          [1 + G^{(R)}_{\alpha} T^{\cm}_{\alpha R}] G^{(R)}_{\alpha},
          \label{eq:GRtU0}
\end{eqnarray}
where the reduced breakup operator $\tilde{U}^{(R)}_{0\alpha}$ may be calculated
through quadrature
\begin{equation} \label{eq:tU0}
\tilde{U}^{(R)}_{0\alpha} = {} 
    G_{\alpha R}^{-1} + v_{\alpha}  + u_\alpha
    + \sum_\gamma (v_\gamma + u_\gamma) G_{\gamma R} \tilde{U}^{(R)}_{\gamma\alpha},
\end{equation}
from the corresponding reduced operator $\tilde{U}^{(R)}_{\beta\alpha}(Z)$
of elastic/rearrangement scattering.
On-shell, the reduced operator
$\tilde{U}^{(R)}_{0\alpha}(Z)$ calculated between screened Coulomb
distorted initial and final states is of finite range, though the
two contributions in Eq.~(\ref{eq:tU0}) have slightly different range properties
as discussed in Ref.~\cite{deltuva:05d}.
The relation between the full and reduced breakup operators is
  \begin{equation}\label{eq:U0t}
      U^{(R)}_{0\alpha} = {}  (1 + T_{\rho R} G_{0})
      \tilde{U}^{(R)}_{0\alpha} (1 + G^{(R)}_{\alpha} T^{\cm}_{\alpha R}).
  \end{equation}
In the full breakup operator $U^{(R)}_{0 \alpha}(Z)$
the external distortions show up in screened
Coulomb waves generated by $(1 + G^{(R)}_{\alpha} T^{\cm}_{\alpha R})$
in the initial state and by $(1 + T_{\rho R} G_{0})$ in the final
state; both wave functions do not have proper limits as $R \to \infty$.
Therefore  the full breakup transition amplitude  in the case of
the unscreened Coulomb potential is obtained via the renormalization 
of the on-shell breakup transition matrix $ U^{(R)}_{0 \alpha}$
in the infinite $R$ limit
\begin{equation}
      \langle \mathbf{p}' \mathbf{q}'  |  U^{(C)}_{0 \alpha}
      |b_\alpha \mathbf{q} \rangle  = 
      \lim_{R \to \infty} [ z^{-\frac{1}{2}}_R(p')
      \langle \mathbf{p}' \mathbf{q}' | 
      U^{(R)}_{0 \alpha}
      |b_\alpha \mathbf{q} \rangle Z_{\alpha R}^{-\frac{1}{2}}(q) ],  \quad
\label{eq:UC1a}
 \end{equation}
where  $\mathbf{p}'$ is the relative momentum between the charged particles 
in the  final state, 
$\mathbf{q}'$ the corresponding particle-pair  relative momentum, and 
\begin{equation}  \label{eq:phiRp}
z_R(p') = e^{-2i\kappa(p')[\ln{(2pR)} - C/n]}
  \end{equation}
the final-state renormalization factor.
The limit in Eq.~(\ref{eq:UC1a}) has to be performed numerically,
but, due to the short-range nature of the breakup operator,
the convergence with the increasing screening radius $R$
is fast and the limit is reached with sufficient accuracy at
finite $R$.

\subsection{Practical realization}

\begin{figure}[!]
\centering
\includegraphics[width=0.85\columnwidth]{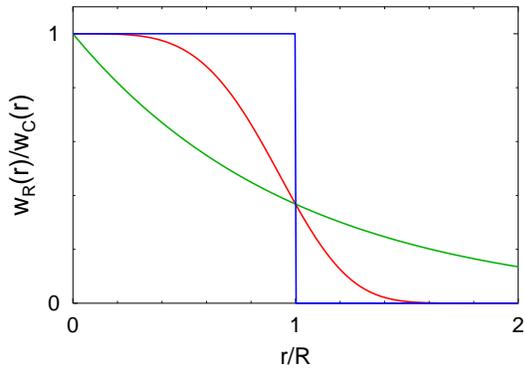}
\caption{
Screening function $w_R(r)/w_C(r)$ as function of the distance $r$ between charged particles
for characteristic values of the parameter $n$ in Eq.~(\ref{eq:wr}):
$n=1$ (green curve) corresponds to Yukawa screening,
$n=4$ (red curve) is our standard choice, 
and $n \to \infty$ (blue curve) corresponds to a sharp cutoff.}
\label{fig:wr}       
\end{figure}

\begin{figure}[!]
\centering
\includegraphics[width=0.9\columnwidth]{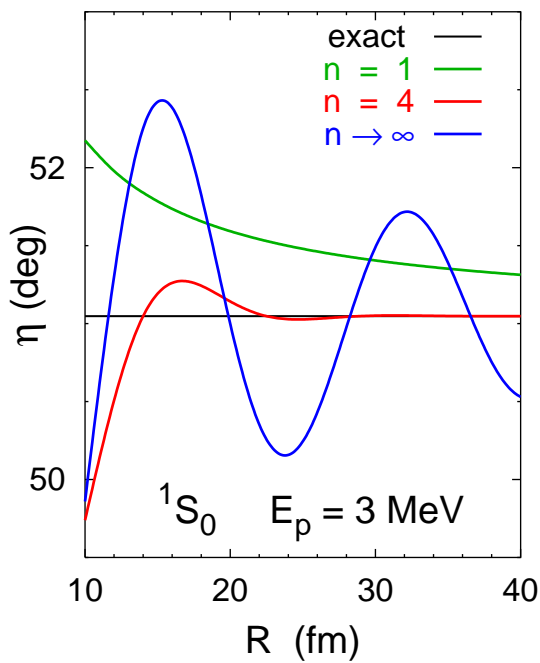}
\caption{
Convergence of the ${}^1S_0$ $pp$ phase shift $\eta$ with screening radius $R$
for 3 MeV proton lab energy.
Results obtained with $n=1$ (green curve), $n=4$ (red curve),
and  $n\to \infty$ (blue curve) screening functions
are compared with exact value given by black line.}
\label{fig:pp}       
\end{figure}

To calculate the short-range part of the elastic, rearrangement, and breakup
scattering amplitudes (\ref{eq:UC2}) and (\ref{eq:UC1a}) we 
solve standard scattering  equations (\ref{eq:Uba}), (\ref{eq:Tcm}),
 and (\ref{eq:U0a}) at finite  Coulomb screening radius $R$ 
using the momentum-space partial-wave representation as described in detail in 
Refs. \cite{chmielewski:03a,deltuva:03a,deltuva:phd}.
We have to make sure that $R$ is large enough to achieve
(after renormalization) the $R$-independence of the results up to a desired accuracy.
However, those $R$ values are larger
than the range of the nuclear interaction resulting in a slower convergence
of the partial-wave expansion. As we found in Ref.~\cite{deltuva:05a},
the practical success of the screening and renormalization method depends strongly
on the choice of the screening function, i.e., on the choice of the exponent $n$
in Eq.~(\ref{eq:wr}). One of the essential  differences compared to 
previous works \cite{alt:94a,alt:02a,berthold:90a,alt:98a}
is that we use a sharper screening than the Yukawa screening $(n=1)$.
We want to ensure that the
screened Coulomb potential $w_R$ approximates well the true Coulomb one
$w_C$ for distances $r<R$  and simultaneously vanishes rapidly for $r>R$,
providing a comparatively fast convergence of the partial-wave expansion.
 However, the sharp cutoff  $(n \to \infty)$
yields an unpleasant oscillatory behavior in the momentum-space representation,
leading to convergence problems.
We find values $3 \le n \le 8$ to provide a sufficiently smooth, but at the
same time a sufficiently rapid screening around $r=R$. The screening functions
for different $n$ values are compared in Fig.~\ref{fig:wr},
showing that the present choice $n=4$ includes much more of the exact
Coulomb potential at short distances than the Yukawa screening used previously.
For example, Yukawa screening requires a screening radius of $R = 1280$ fm in order to
approximate true Coulomb at relative distance $r = 5$ fm as well as the present choice does
with $R = 20$ fm. As shown in Fig.~\ref{fig:pp} for the proton-proton ($pp$)
${}¹S_0$ phase shift, the convergence with the screening radius for
$n=4$ is much faster than for Yukawa screening or sharp cutoff.
Furthermore, due to the much shorter range of the $n=4$ screening compared to the 
Yukawa screening, the quasisingularities of the screened Coulomb potential are
far less pronounced in the $n=4$ case.
With our optimal choice  $3 \le n \le 8$ the convergence of the
partial-wave expansion, though being slower than for the nuclear interaction alone,
can still be achieved; alternatively, a perturbative approach for higher two-particle
partial waves \cite{deltuva:03b} that is computationally less demanding but, 
nevertheless, highly reliable as proved in Ref.~\cite{deltuva:06a}, can be used as well.

\subsection{Renormalization in proton-deuteron scattering}

\begin{figure*}[!]
\begin{center}
\includegraphics[width=0.99\columnwidth]{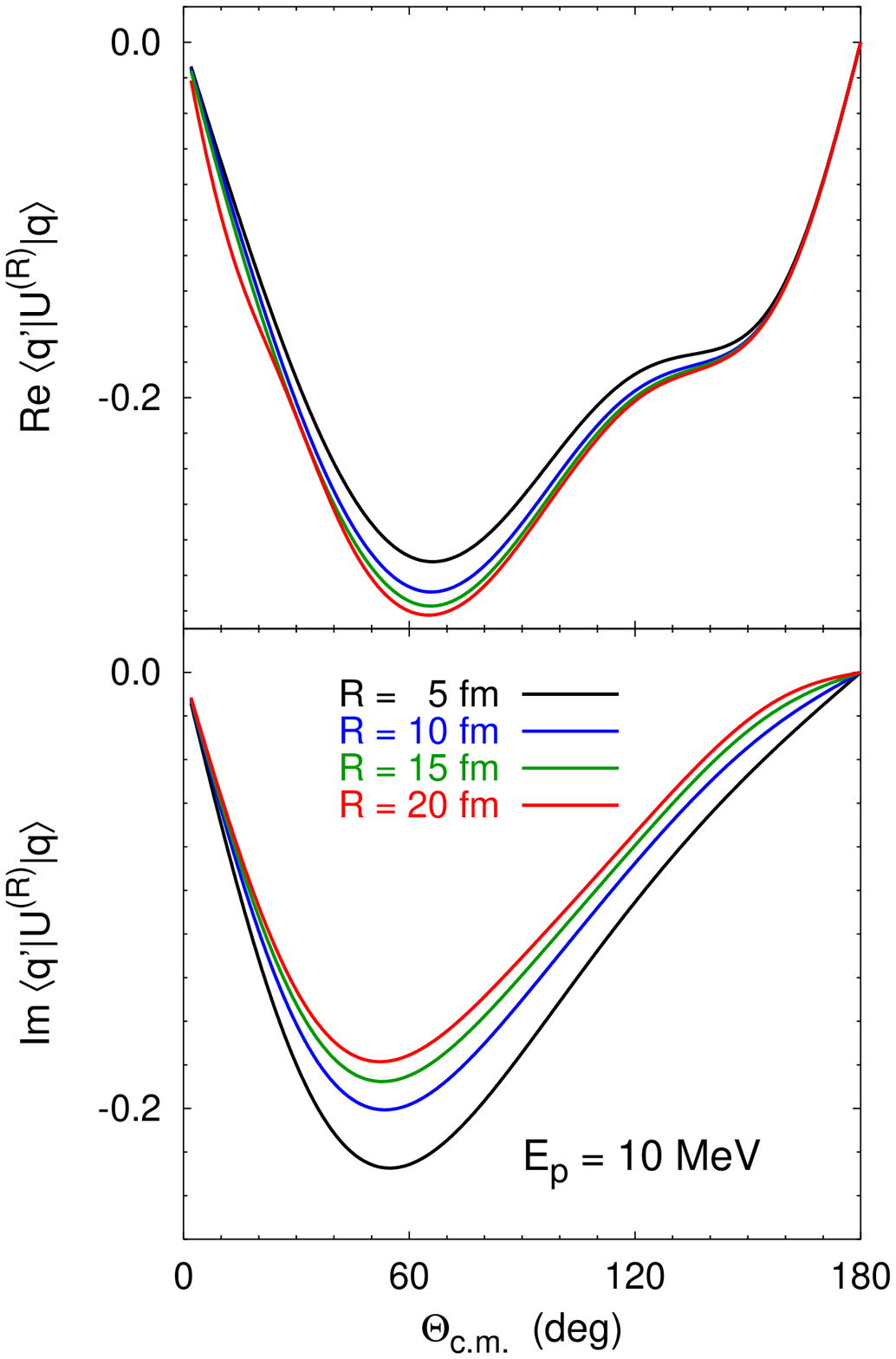} \quad
\includegraphics[width=0.99\columnwidth]{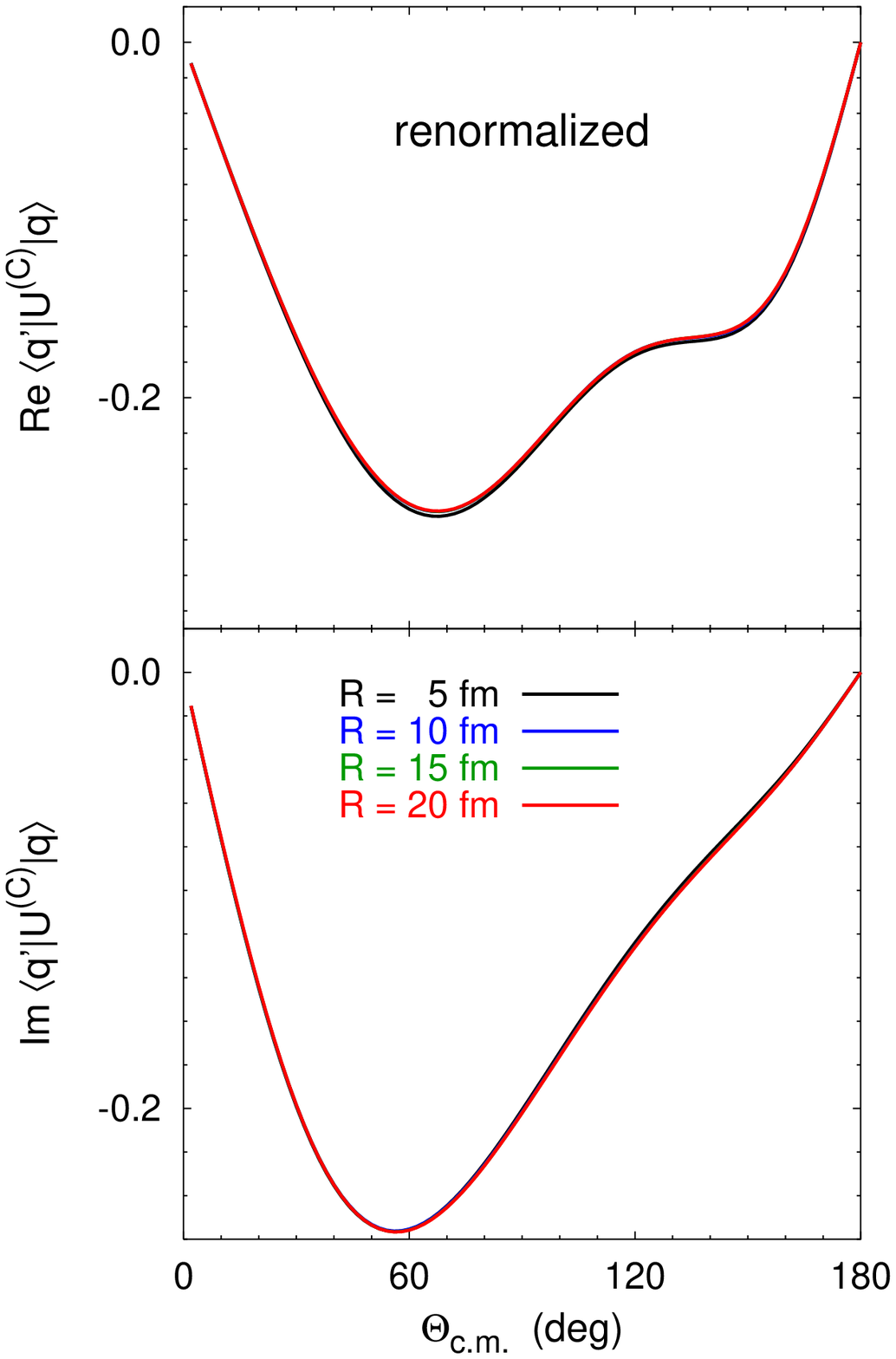}
\end{center}
\caption{\label{fig:URZs} 
Real and imaginary parts of
nonrenormalized (left side) and renormalized (right side) $p$-$d$ elastic
scattering amplitudes (in arbitrary units) at $E_p = 10$ MeV
shown as functions of the c.m. scattering angle.
The initial and final particle spin projection
quantum numbers are $m_p=\frac12$, $m'_p=-\frac12$, and  $m_d = m'_d = 1$.
Results for the  CD Bonn + $\Delta$ two-baryon potential
 obtained with the screening radius
$R= 5$~fm (black curves), 10~fm (blue curves),
15~fm (green curves), and 20~fm (red curves) are compared.}
\end{figure*}

For the nucleon-deuteron scattering it is convenient to
 consider nucleons as identical particles and use the isospin formalism.
Since the isospin conservation is violated by the Coulomb force, 
both total $3N$ isospin $\mathcal{T}=\frac12$ and $\frac32$
states have to be included.
 The symmetrized transition operator for elastic scattering
is the solution of the symmetrized AGS integral equation \cite{deltuva:09e}
\begin{eqnarray} \nonumber
U^{(R)} &=& {}  PG_0^{-1} + (1+P)u + PT^{(R)}G_0U^{(R)} \\ & &
 + (1+P)uG_0(1+T^{(R)}G_0)U^{(R)}.
\label{eq:AGSs}
\end{eqnarray}
We omit the spectator index that is not needed anymore. The basis states are
antisymmetric in the pair only, the full antisymmetry is ensured
by $P = P_{12}P_{23} + P_{13}P_{23}$ where $P_{\alpha \beta}$ is  the
permutation operator of particles $\alpha$ and  $\beta$.
The breakup operator is then obtained from the quadrature
\begin{equation}
U^{(R)}_0 = (1+P)[G_0^{-1} +u +T^{(R)}G_0U^{(R)} + uG_0(1+T^{(R)}G_0)U^{(R)}].
\end{equation}
The amplitudes for $p$-$d$ elastic scattering and breakup referring to
 unscreened Coulomb are obtained after the renormalization of the
corresponding on-shell matrix elements of the symmetrized transition operators, i.e.,
\begin{eqnarray} \nonumber 
\langle \mathbf{q}' |U^{(C)} | \mathbf{q} \rangle  &= {} & \nonumber
    \langle \mathbf{q}' |T^{\cm}_C |\mathbf{q} \rangle \\ & &    \label{eq:UC}
+    \lim_{R \to \infty}     [ Z_{R}^{-1}(q)
\langle \mathbf{q}' | (U^{(R)}- T^{\cm}_R) |\mathbf{q} \rangle ], \quad \\ \label{eq:UC0}
\langle \mathbf{p}'\mathbf{q}' |U_0^{(C)} |\mathbf{q} \rangle  &= {} &
\lim_{R \to \infty} [ z_R^{-\frac12}(p') 
\langle \mathbf{p}'\mathbf{q}' |U_0^{(R)} |\mathbf{q} \rangle
Z_{R}^{-\frac12}(q) ].
\end{eqnarray}

\begin{figure*}[!]
\begin{center}
\includegraphics[width=0.99\columnwidth]{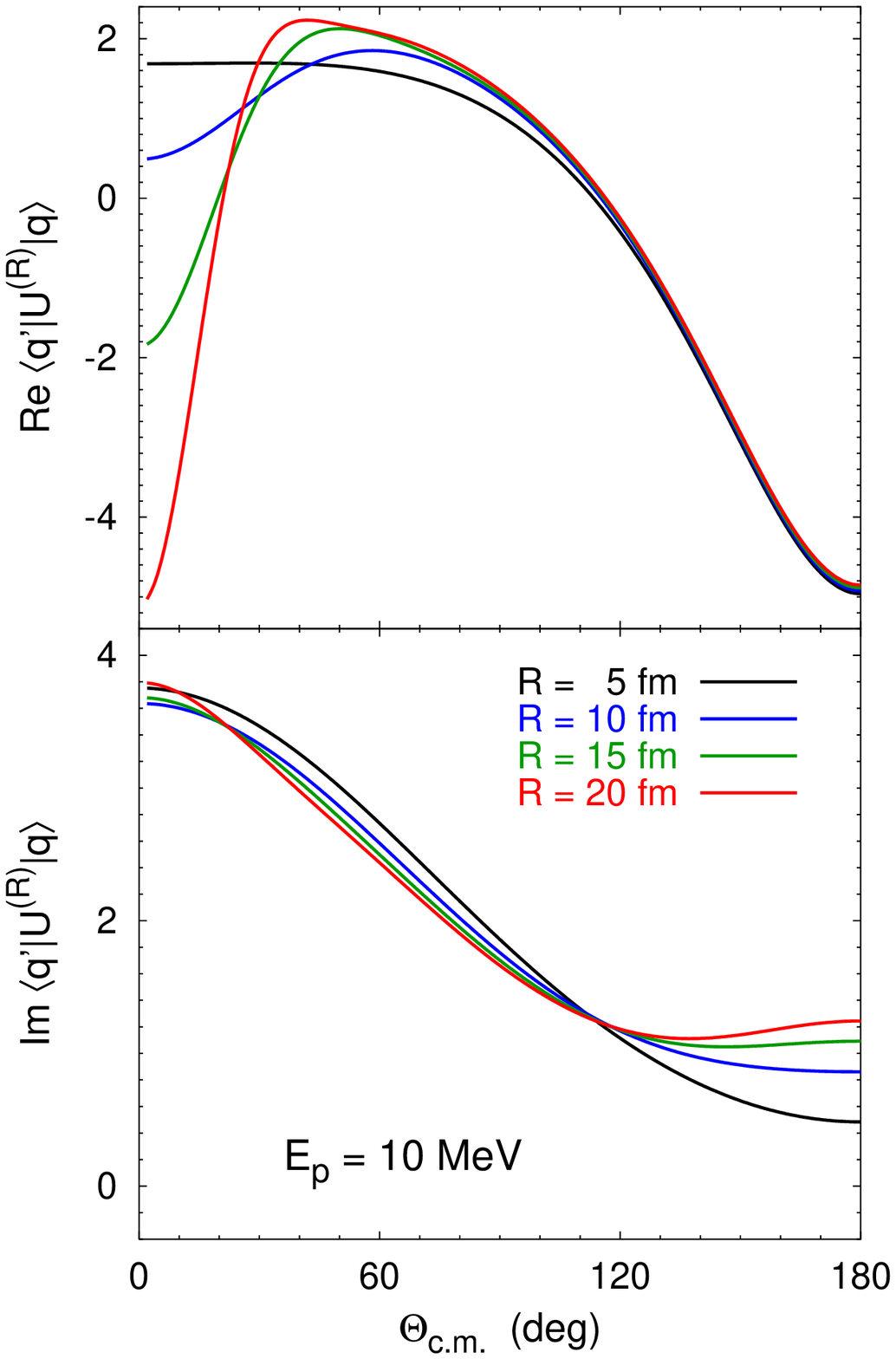} \quad
\includegraphics[width=0.99\columnwidth]{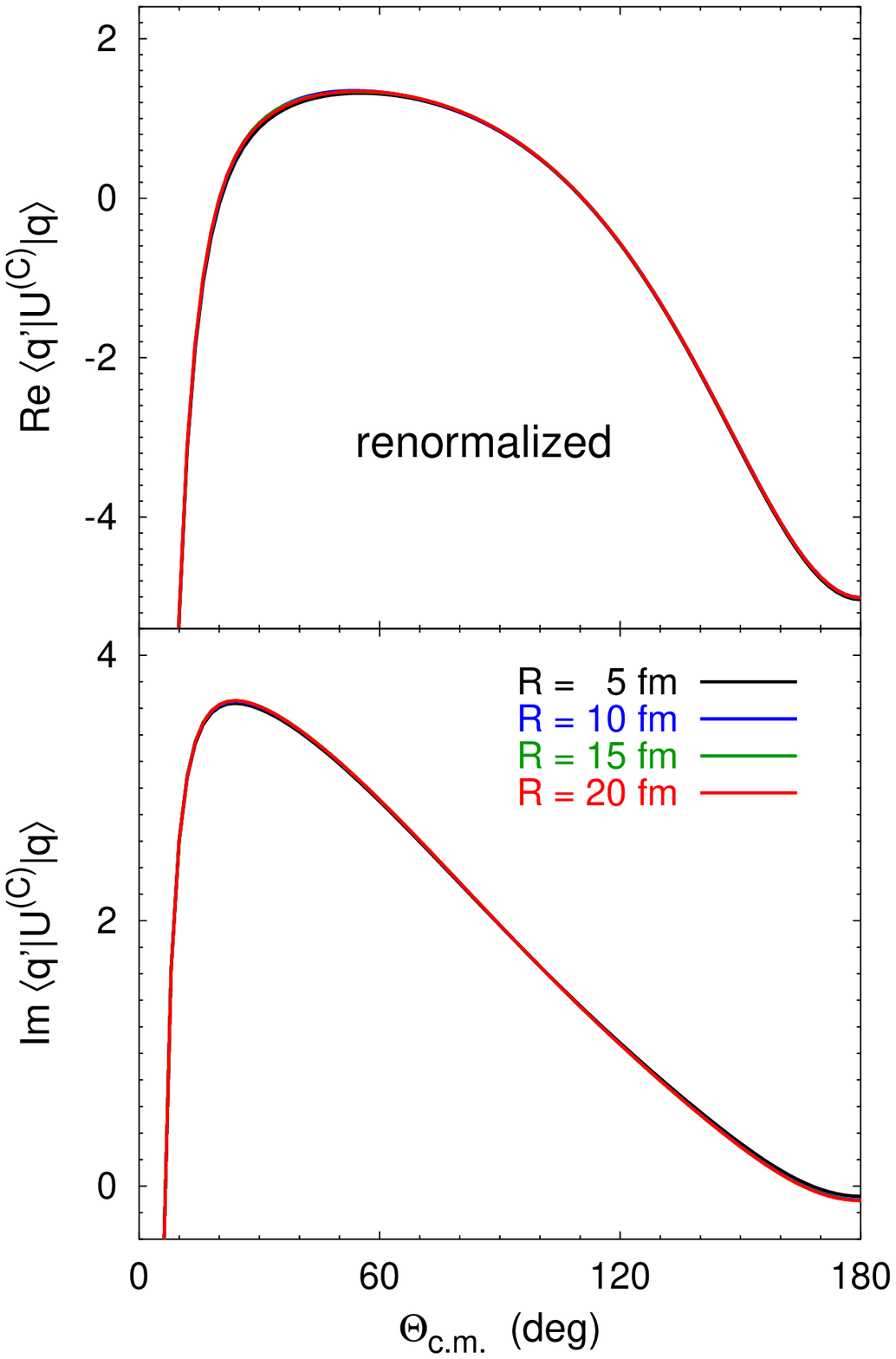}
\end{center}
\caption{\label{fig:URZl} 
Real and imaginary parts of
nonrenormalized (left side) and renormalized (right side) $p$-$d$ elastic
scattering amplitudes (in arbitrary units) at $E_p = 10$ MeV
shown as functions of the c.m. scattering angle.
The initial and final particle spin projection
quantum numbers are $m_p=m'_p=\frac12$,  and  $m_d = m'_d = 1$.
Curves as in Fig.~\ref{fig:URZs}.}
\end{figure*}

\begin{figure*}[!]
\begin{center}
\includegraphics[width=0.99\columnwidth]{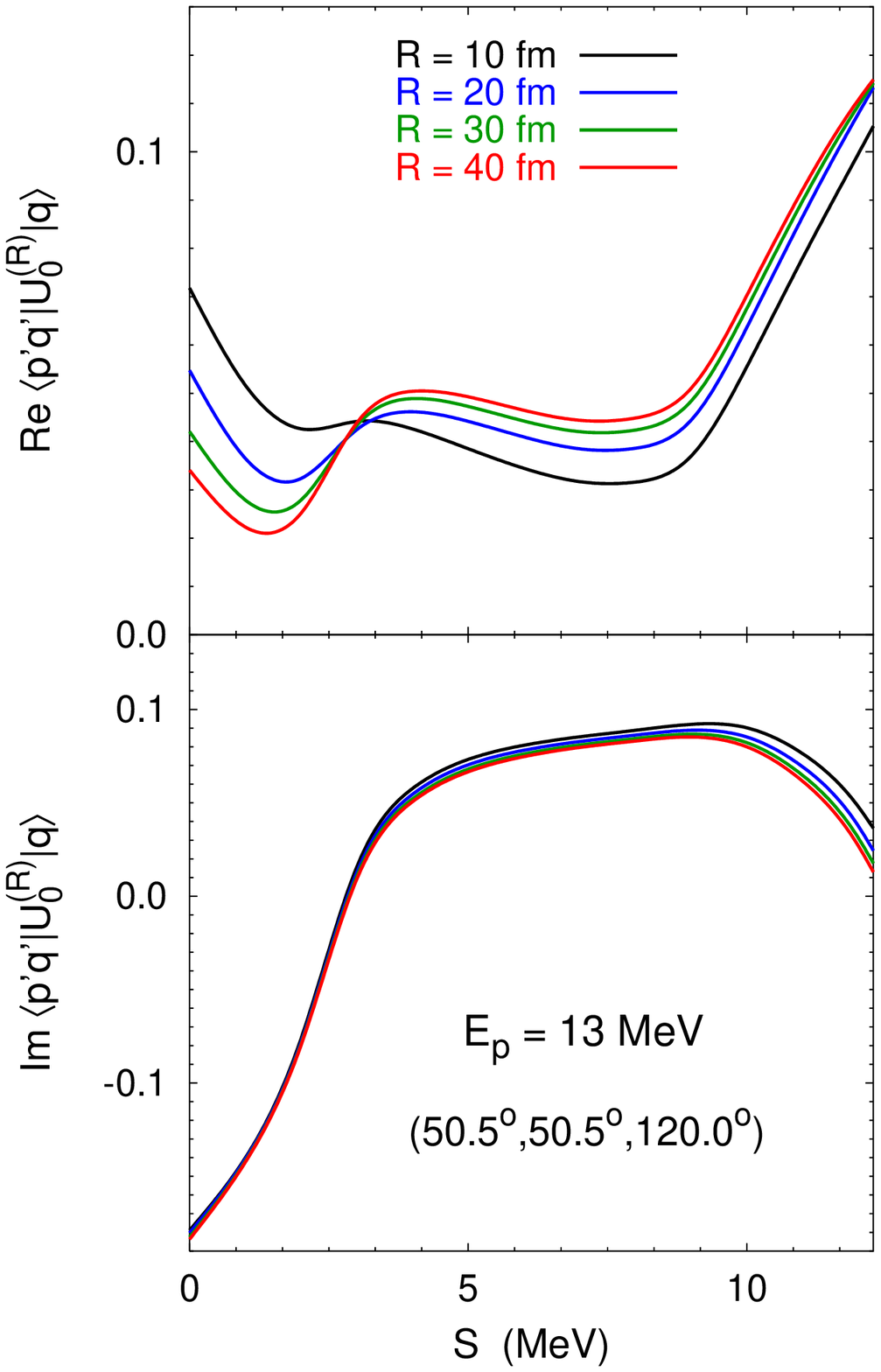} \quad
\includegraphics[width=0.99\columnwidth]{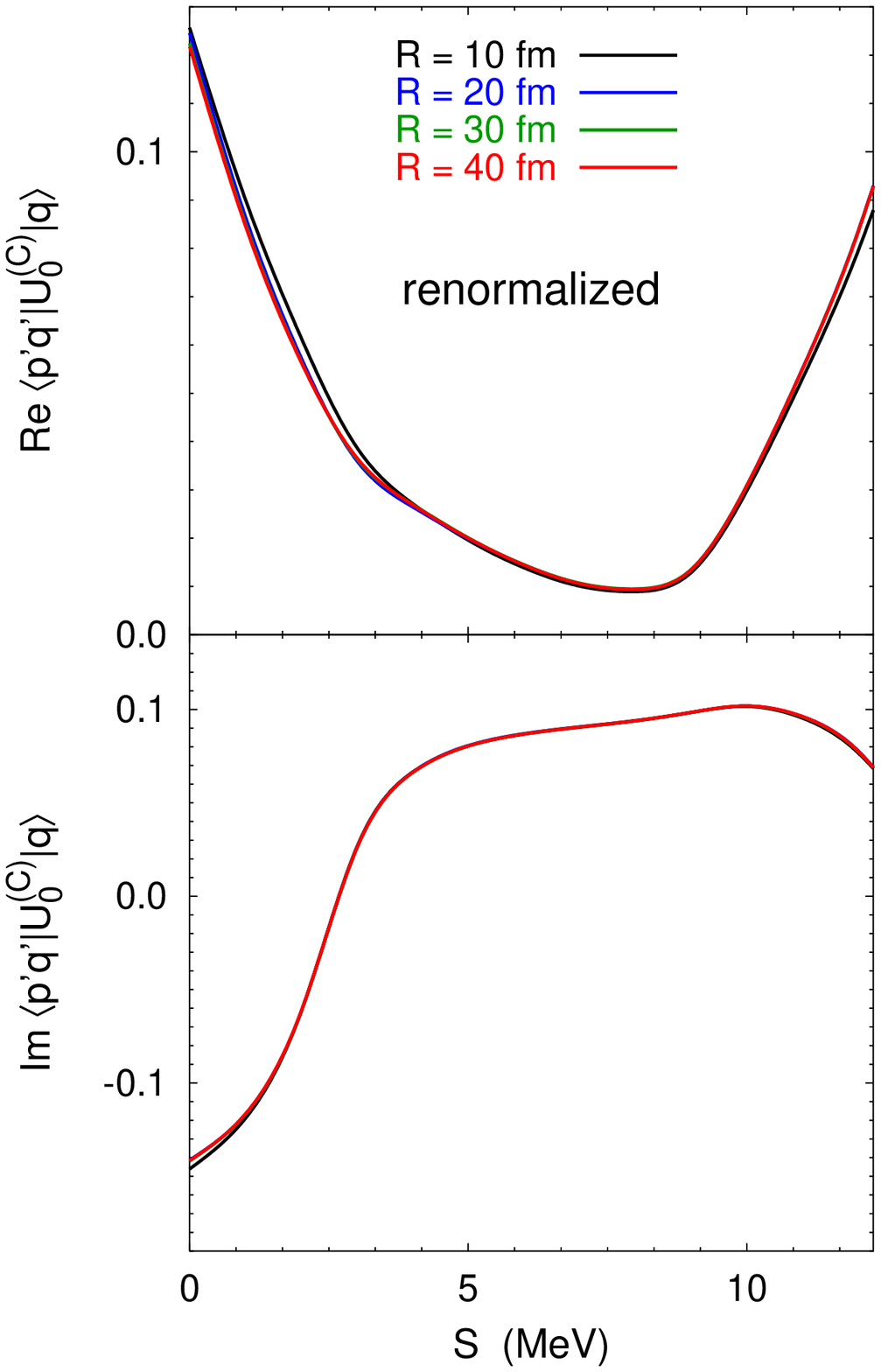}
\end{center}
\caption{\label{fig:U0RZ} 
Real and imaginary parts of nonrenormalized (left side) and renormalized by 
$\exp\{i[\kappa(p)+\mathcal{K}(q)]\ln R\}$
(right side) $p$-$d$ breakup
amplitudes (in arbitrary units) at $E_p = 13$ MeV
in the space star configuration
shown as functions of the arclength $S$ along the kinematical curve.
Curves as in Fig.~\ref{fig:URZs}.}
\end{figure*}

However, a recent work \cite{witala:09a}
on an alternative Coulomb treatment in $pd$ scattering
proposed a different renormalization prescription:
 $p$-$d$ elastic scattering amplitude calculated with screened Coulomb
does not need renormalization at all, i.e., the limit
$\lim_{R \to \infty} \langle \mathbf{q}' |U^{(R)}|\mathbf{q} \rangle$ should exist;
the renormalization for the breakup amplitude is needed but it is different
from ours given in Eq.~(\ref{eq:UC0}).
However, numerical results of Ref.~\cite{witala:09a}
involve approximations that are not well under control. First, the screened
Coulomb proton-proton transition matrix is approximated by the screened
Coulomb potential thereby replacing a complex quantity by a real one.
Second, the contributions \\
$\langle \mathbf{q}' |P T_R P G_0 T^{(R)}G_0U^{(R)} | \mathbf{q} \rangle$ and \\
$\langle \mathbf{p}'\mathbf{q}' |(1+P) T_R P G_0 T^{(R)}G_0U^{(R)} | \mathbf{q} \rangle$ 
to the elastic scattering and breakup amplitudes, respectively, are neglected;
this may have even more serious consequences since
the above  terms among others include screened Coulomb contributions that are of the
leading (first) order. 
Indeed, under the approximations of Ref.~\cite{witala:09a} the $R$-dependence of the
resulting screened Coulomb amplitude for $p$-$d$ elastic scattering 
is weak and the $R \to \infty$ limit of renormalized amplitude
in Eq.~(\ref{eq:UC}) does not exists.
However, the situation changes completely when the scattering amplitudes
are calculated exactly as in our work.
In  Figs. \ref{fig:URZs} and \ref{fig:URZl}
we study the dependence on the screening radius $R$ for selected components of the
nonrenormalized and renormalized $p$-$d$ elastic scattering amplitudes
$\langle \mathbf{q}' | U^{(R)} |\mathbf{q} \rangle $ and 
$\langle \mathbf{q}' | U^{(C)} |\mathbf{q} \rangle $.
The CD Bonn + $\Delta$ two-baryon potential \cite{deltuva:03c} is taken
as the hadronic interaction and the proton lab energy $E_p = 10$ MeV.
In Fig.~\ref{fig:URZs} we show spin-nondiagonal amplitudes that have
only the Coulomb-distorted short-range part.
The nonrenormalized amplitude
$\langle \mathbf{q}' | U^{(R)} |\mathbf{q} \rangle $ shows a clear $R$-dependence;
in fact, its absolute value becomes $R$-independent but the phase changes like
 $\ln R$. In contrast,
the renormalized amplitude $\langle \mathbf{q}' | U^{(C)} |\mathbf{q} \rangle $,
within the accuracy of the plot, becomes independent of $R$ for $R \ge 10$ fm.
The  spin-diagonal amplitude  having also a long-range part shows, as expected,
very strong $R$-dependence at small scattering angles before renormalization
but  becomes $R$-independent after renormalization as well
as Fig.~\ref{fig:URZl} demonstrates. In Fig. \ref{fig:U0RZ} we show the corresponding
results for $p$-$d$ breakup at $E_p = 13$ MeV in the space star configuration.
For breakup reaction the renormalized amplitude 
$\langle \mathbf{p}'\mathbf{q}' | U^{(C)}_0 |\mathbf{q} \rangle $ 
converges with the screening radius somehow slower than for elastic scattering,
it becomes independent of $R$ for $R \ge 20$ fm whereas 
the nonrenormalized amplitude
$\langle \mathbf{p}'\mathbf{q}' | U^{(R)}_0 |\mathbf{q} \rangle $ 
shows a clear $R$-dependence. Furthermore, although the breakup amplitude
(without the three-nucleon force) can be decomposed into three terms 
$ T^{(R)}_\gamma G_0 U^{(R)}_{\gamma \alpha}$, separately none of them has an $R \to \infty$
limit with or without renormalization in contrast to the conjecture of 
 Ref.~\cite{witala:09a}; infinite $R$ limit exists only for renormalized full  
breakup amplitude. 
Thus, fully converged numerical results without uncontrolled approximations
clearly support the standard screening and renormalization theory as given in
Eqs. (\ref{eq:UC}) and (\ref{eq:UC0}) and not the one of Ref.~\cite{witala:09a}.

The convergence of the observables with the screening radius $R$ used to calculate the 
Coulomb-distorted short-range part of the amplitudes  is the internal criterion for 
the reliability of our method.
Numerous examples can be found in 
Refs. \cite{deltuva:05a,deltuva:05d,deltuva:06b,deltuva:fb18}.
In most cases the convergence is impressively fast;
the screening radius $R = 10 $ to 30 fm is sufficient.
The exceptions requiring larger screening radii are the 
observables at very low energies and the breakup
differential cross section in kinematical situations characterized 
by very low relative energy $E_{\mathrm{rel}}$ between the two
charged particles, e.g., $p$-$d$ breakup or photodisintegration of ${}^3\mathrm{He}$
close to the $pp$ final-state interaction  ($pp$-FSI) regime \cite{deltuva:05d}.
The slow convergence under those conditions is not surprising, since the 
renormalization factor itself as well as the Coulomb parameter
become ill-defined, indicating that the  screening and renormalization
procedure cannot be applied at $E_{\mathrm{rel}}=0$.
Therefore an extrapolation has to be used to calculate
the observables at $E_{\mathrm{rel}}=0$, which works pretty well when
the observables vary smoothly with $E_{\mathrm{rel}}$ which is the case
in three-body breakup reactions.
Furthermore, the results for $p$-$d$ elastic scattering obtained by the 
present technique were compared \cite{deltuva:05b} with
those of Ref.~\cite{kievsky:01a} obtained from the variational solution
of the three-nucleon Schr\"odinger equation in configuration space
with the inclusion of an \emph{unscreened} Coulomb potential
between the protons and imposing  the proper Coulomb boundary
conditions explicitly. Good agreement over a wide energy range
was found indicating that both techniques for including the Coulomb interaction
are reliable. At very low energies the coordinate-space treatments remain 
favored since there the method of screening and renormalization 
converges slowly and therefore becomes technically too demanding, 
but at higher energies and for three-body breakup reactions it is more 
efficient. 

\section{Results}

In this section we  present selected results for reactions in 
various three- and four-body nuclear systems. 

\subsection{Proton-deuteron scattering}

\begin{figure*}[!]
\begin{center}
\includegraphics[width=1.99\columnwidth]{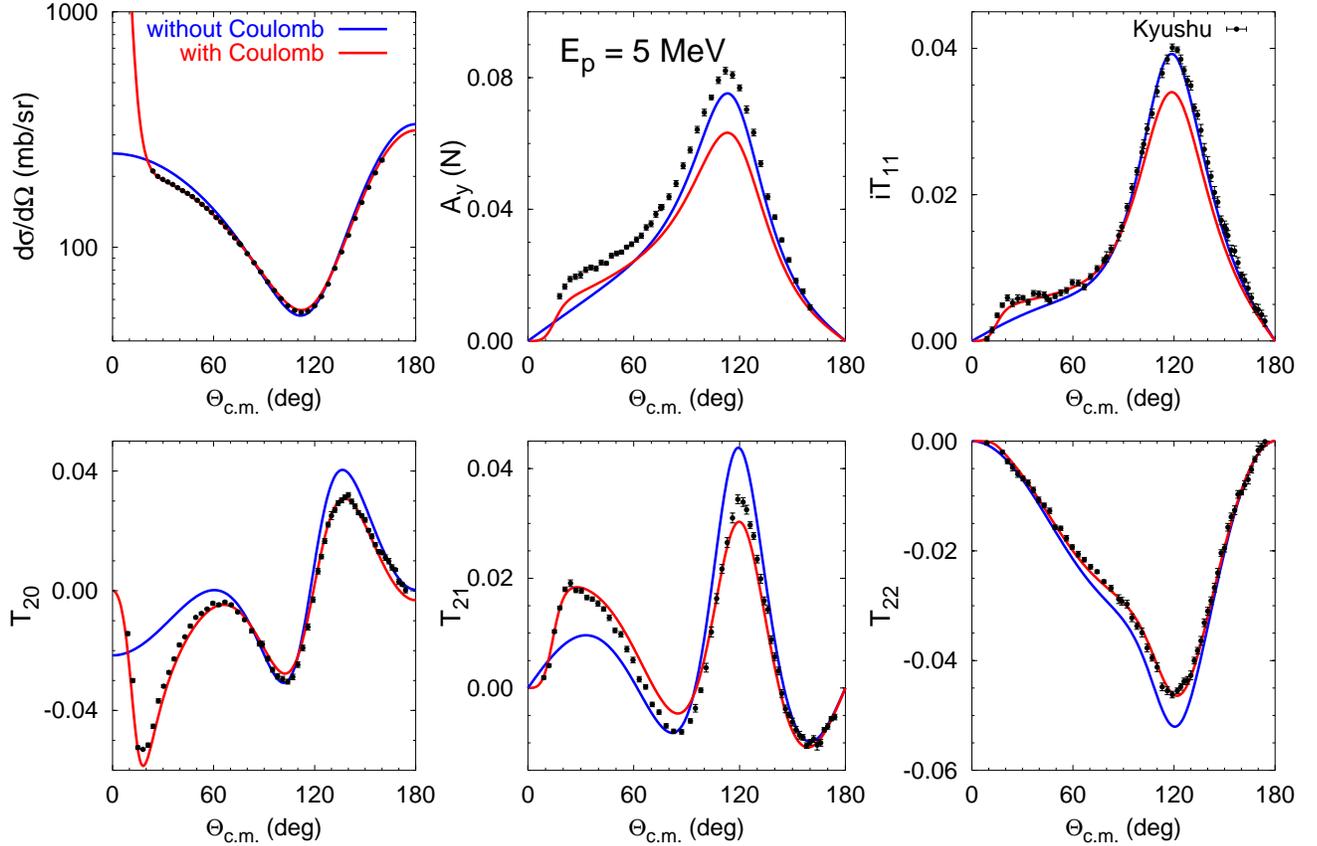}
\end{center}
\caption{\label{fig:pd5}
Differential cross section and analyzing powers for $p$-$d$ elastic scattering 
at 5~MeV proton lab energy as functions of the c.m. scattering angle. 
Results including  the Coulomb interaction
(red curves) are compared to results without Coulomb (blue curves).
Hadronic potential is CD Bonn + $\Delta$.
The experimental data are from Ref.~\cite{sagara:94a}. } 
\end{figure*}

\begin{figure*}[!]
\begin{center}
\includegraphics[width=1.99\columnwidth]{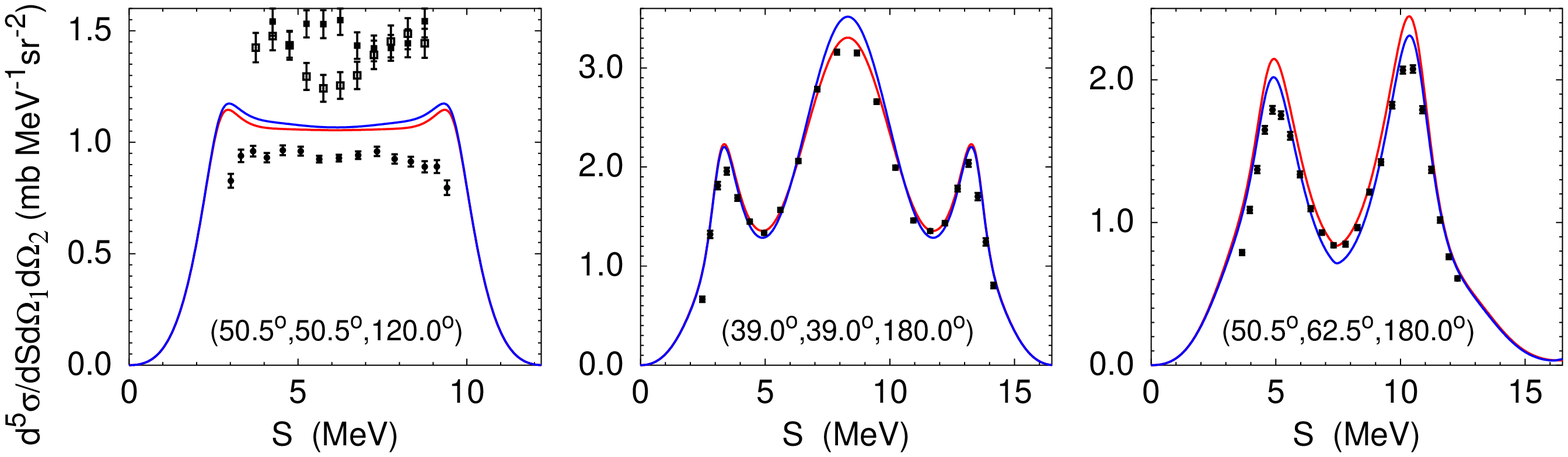}
\end{center}
\caption{\label{fig:pd13} 
Differential cross section for $p$-$d$ breakup at 13~MeV proton lab energy 
in space star (left), quasifree scattering (middle), and
collinear (right) configurations 
as function of the arclength $S$ along the kinematical curve.
Results including  the Coulomb interaction
(red curves) are compared to results without Coulomb (blue curves).
Hadronic interaction model is AV18 + UIX.
The experimental data are from Ref.~\cite{rauprich:91} (full circles).
For the space star configuration  also the $nd$ data from 
Refs.~\cite{strate:89,setze:05a} (open and full squares)
are shown.}
\end{figure*}

Realistic models for the
nuclear interaction are used, e.g., CD Bonn +$\Delta$ two-baryon potential 
\cite{deltuva:03c} yielding an effective three-nucleon force 
or the AV18 two-nucleon potential \cite{wiringa:95a} with the irreducible
Urbana IX (UIX) three-nucleon force \cite{urbana9}.
The Coulomb effect is important in low-energy $p$-$d$ elastic scattering
as Fig. \ref{fig:pd5} demonstrates; the description of the experimental data is quite
satisfactory except for the proton and deuteron vector analyzing powers.
In $p$-$d$ elastic scattering at higher energies the Coulomb effect gets confined to
forward angles.

In Fig.~\ref{fig:pd13} we show results for  $p$-$d$ breakup at 13 MeV proton lab energy.
Although the inclusion of the Coulomb force slightly improves the agreement with
data in the space star configuration, the Coulomb effect is far too small
to reproduce the difference between the experimental $p$-$d$ and $n$-$d$ data and
to resolve the so-called \emph{space star anomaly}.
Slightly larger and beneficial Coulomb effects are seen in quasifree 
scattering (QFS) and collinear configurations.

\begin{figure*}[!]
\begin{center}
\includegraphics[width=1.99\columnwidth]{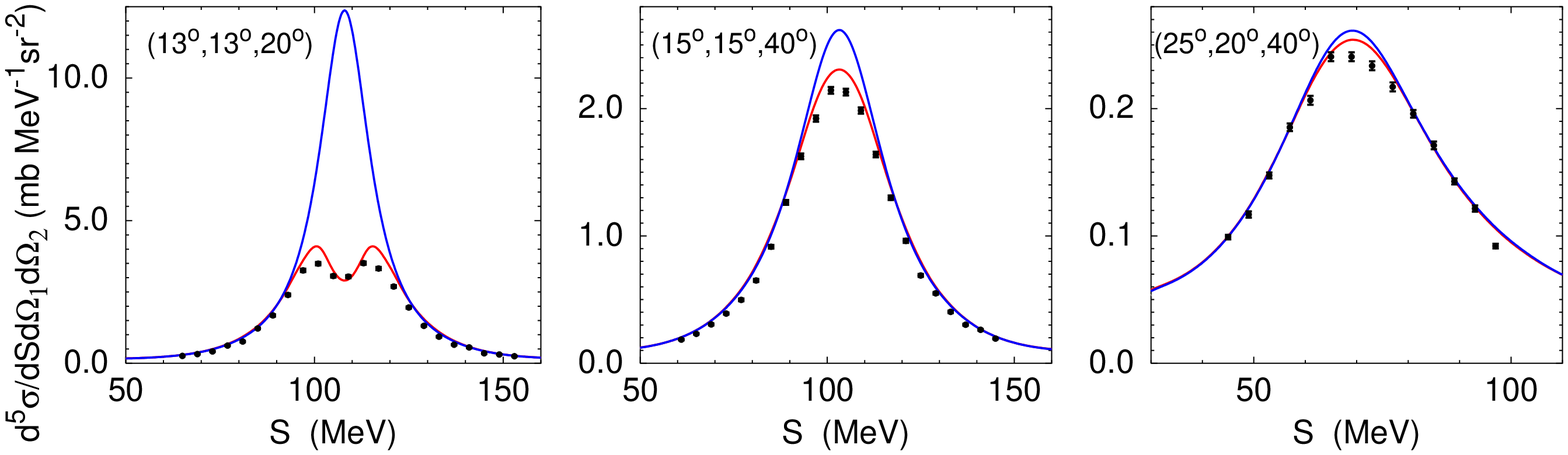}
\end{center}
\caption{\label{fig:dp130_40} 
Differential cross section for $pd$ breakup
at  130~MeV deuteron lab energy in selected kinematical configurations
with small relative azimuthal angle.
Curves as in Fig.~\ref{fig:pd13} and  the 
 experimental data from Refs.~\cite{kistryn:05a,kistryn:06a}.}
\end{figure*}

\begin{figure*}[!]
\begin{center}
\includegraphics[width=1.99\columnwidth]{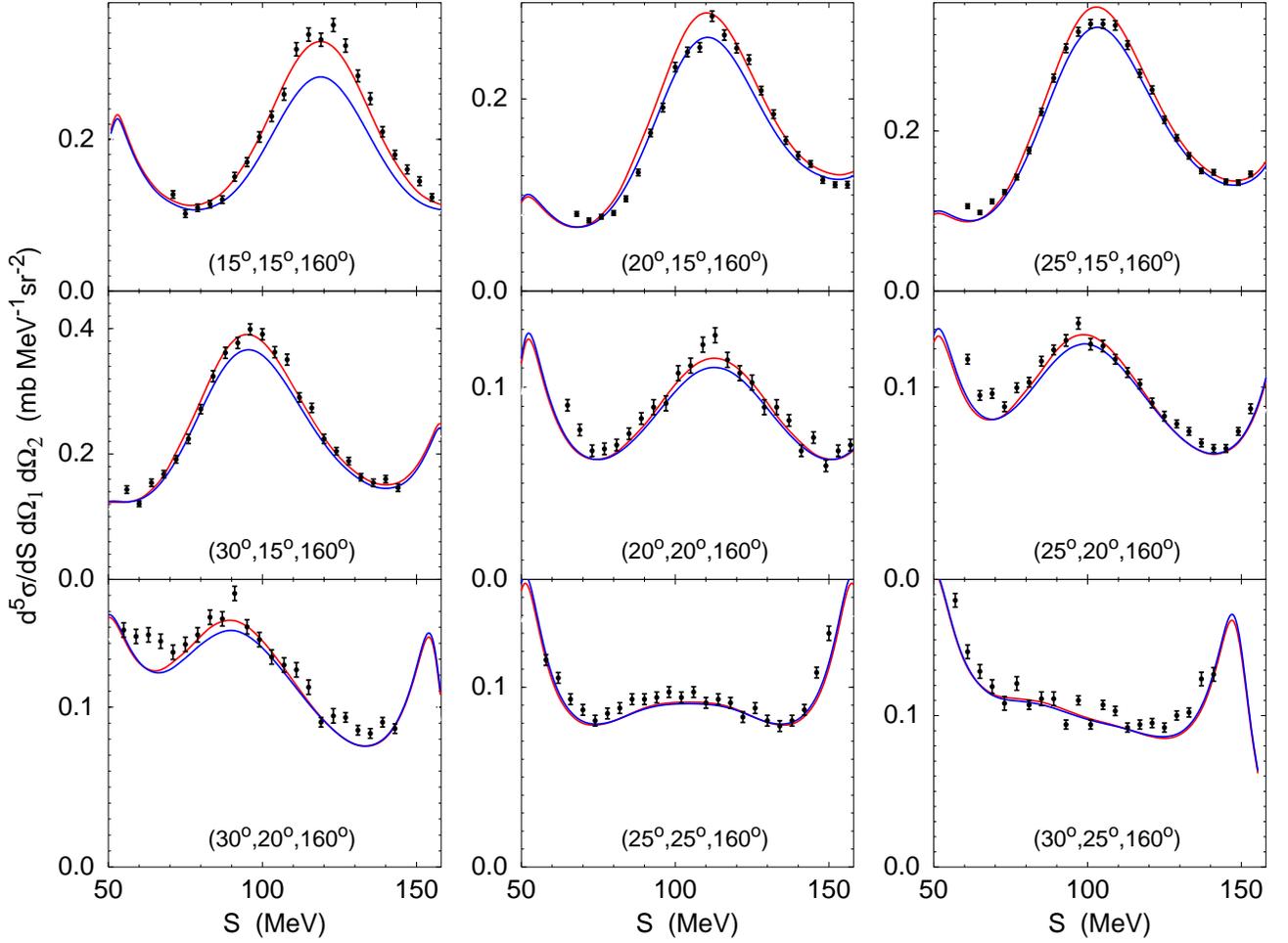}
\end{center}
\caption{\label{fig:dp130_160} 
Differential cross section  for $p$-$d$ breakup
at  130~MeV deuteron lab energy in selected kinematical configurations
with large relative azimuthal angle.
Curves as in Fig.~\ref{fig:pd13} and  the 
 experimental data from Ref.~\cite{kistryn:05a}.}
\end{figure*}

The Coulomb effect in  may become large even at higher energies
when the relative $pp$ energy in the final state is very low as demonstrated 
in Fig.~\ref{fig:dp130_40} for $p$-$d$ breakup at 130 MeV deuteron lab energy.
In there, the Coulomb repulsion is responsible for decreasing the cross section;
the $pp$-FSI peak obtained in the absence of Coulomb may even be converted into a minimum.
However, in some other configurations Coulomb may lead to a moderate increase of the 
differential cross section as shown in Fig.~\ref{fig:dp130_160}.

\subsection{Photodisintegration of ${}^3\mathrm{He}$}

\begin{figure*}[!]
\begin{center}
\includegraphics[width=1.99\columnwidth]{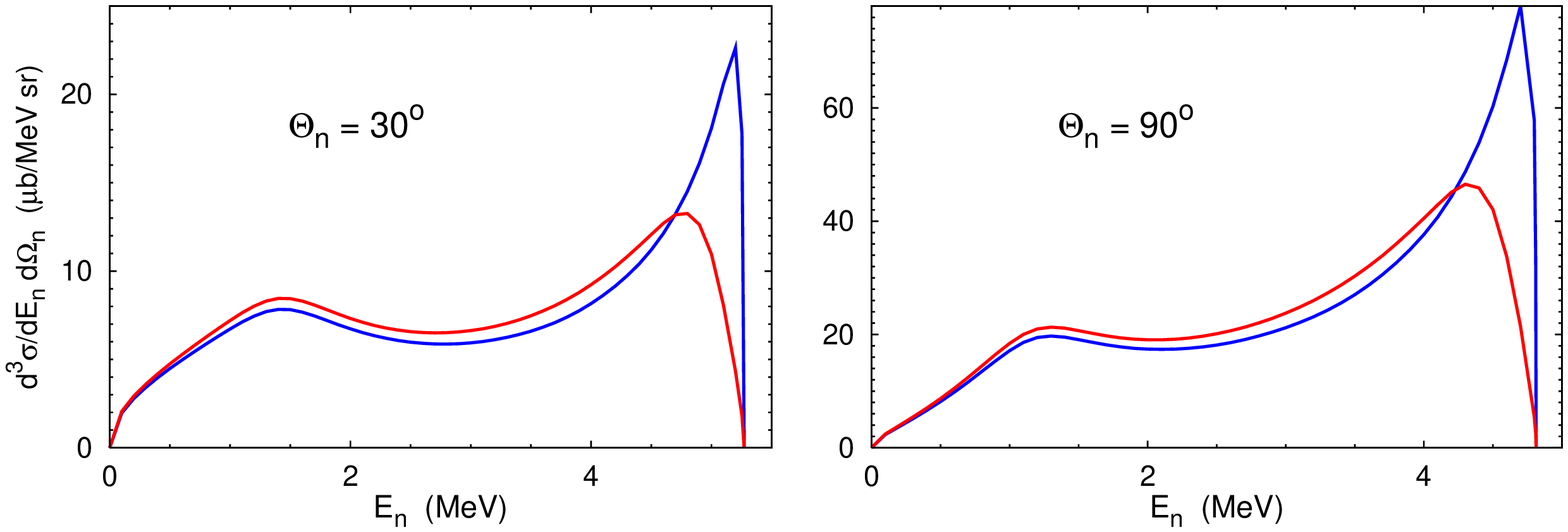}
\includegraphics[width=1.99\columnwidth]{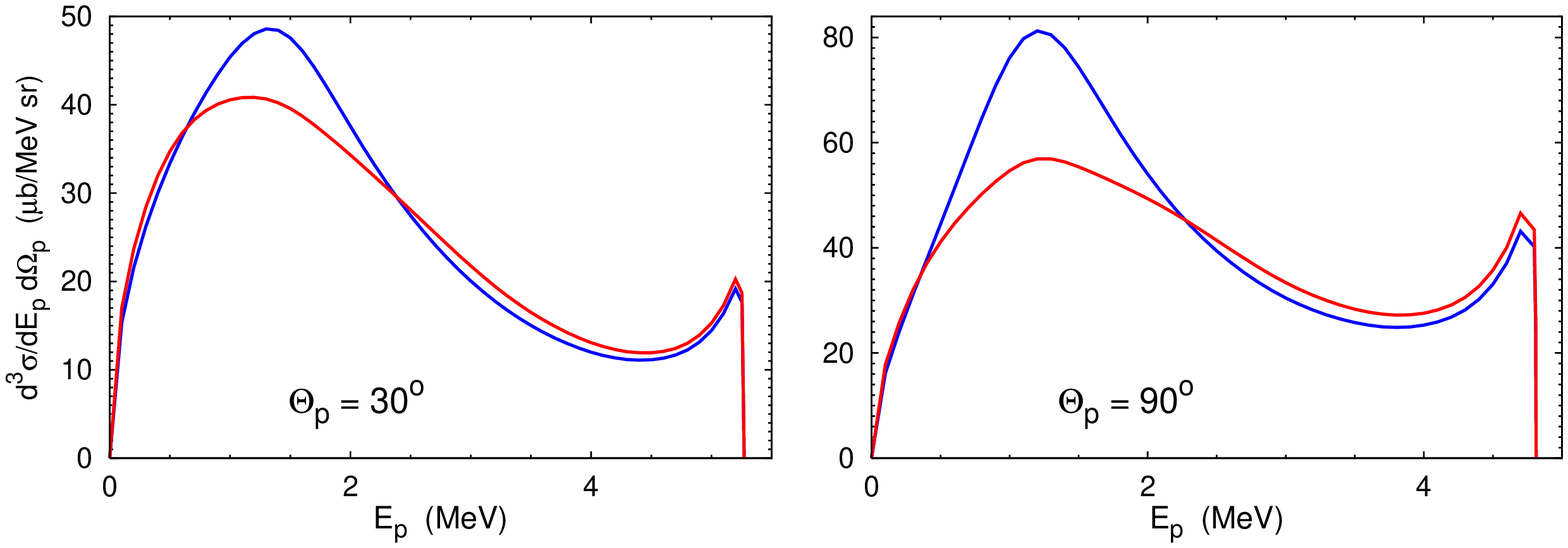}
\end{center}
\caption{\label{fig:g3He} 
The semi-inclusive threefold differential cross section  for 
the ${}^3\mathrm{He}(\gamma,n)pp$ (top) and ${}^3\mathrm{He}(\gamma,p)pn$ (bottom)
reactions at  15~MeV photon lab energy as function of the detected nucleon energy
at $30^{\circ}$ and $90^{\circ}$ nucleon scattering angles.
Curves as in Fig.~\ref{fig:pd5}.}
\end{figure*}

\begin{figure*}[!]
\begin{center}
\includegraphics[width=1.99\columnwidth]{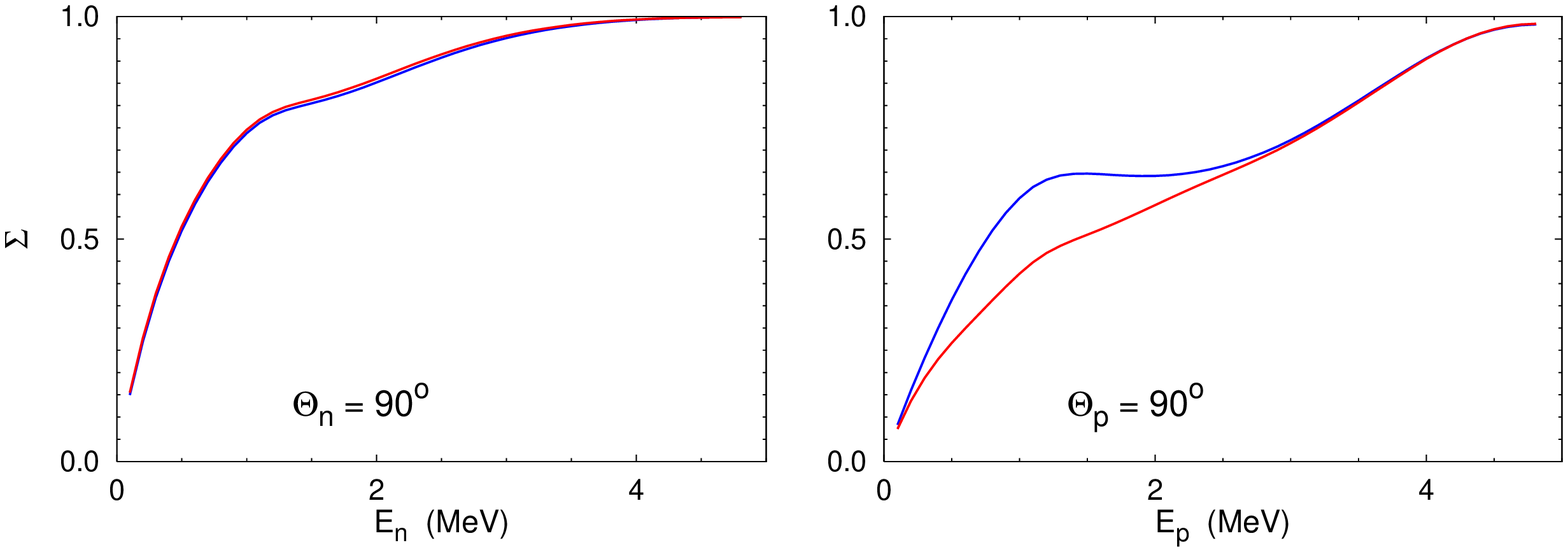}
\end{center}
\caption{\label{fig:g3He-A} 
The semi-inclusive linear photon asymmetry  for 
the ${}^3\mathrm{He}(\gamma,n)pp$ (left) and ${}^3\mathrm{He}(\gamma,p)pn$ (right)
reactions at  15~MeV photon lab energy as function of the detected nucleon energy
at $90^{\circ}$ nucleon scattering angle.
Curves as in Fig.~\ref{fig:pd5}.}
\end{figure*}

The results are obtained with the CD Bonn +$\Delta$ two-baryon potential 
and include effective two- and three-nucleon electromagnetic currents 
mediated by the $\Delta$ isobar \cite{deltuva:04a,deltuva:04b}.
As in $p$-$d$ breakup, the Coulomb effect may be very strong in the three-body 
photodisintegration of ${}^3\mathrm{He}$ close to the $pp$-FSI kinematics.
The effect remains also after partial integration when calculating the
semi-inclusive observables as shown in Figs.~\ref{fig:g3He} and \ref{fig:g3He-A}.
Not only the differential cross section gets affected but also the spin observables
like the linear photon asymmetry  for ${}^3\mathrm{He}(\gamma,p)pn$ reaction.

\subsection{Three-body nuclear reactions}

\begin{figure*}[!]
\begin{center}
\includegraphics[width=1.99\columnwidth]{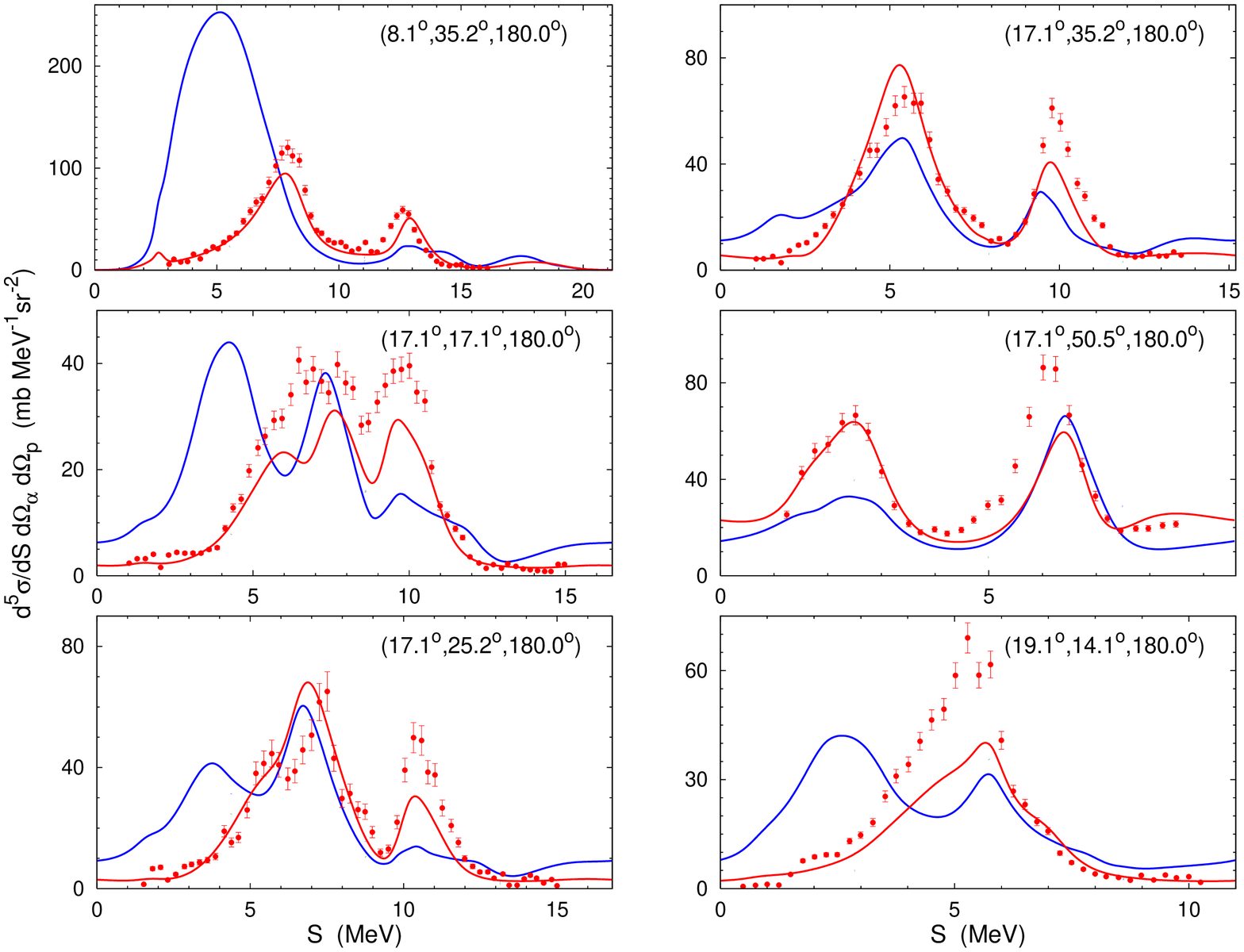}
\end{center}
\caption{\label{fig:ad} 
Differential cross section for $\alpha$-$d$ breakup
at 15~MeV $\alpha$ lab energy in selected kinematical configurations
as function of the arclength $S$ along the kinematical curve.
Results including  the Coulomb interaction
(red curves) are compared to results without Coulomb (blue curves).
The experimental data are from Ref.~\cite{koersner:77}.}
\end{figure*}

\begin{figure*}[!]
\centering
\includegraphics[width=0.98\columnwidth]{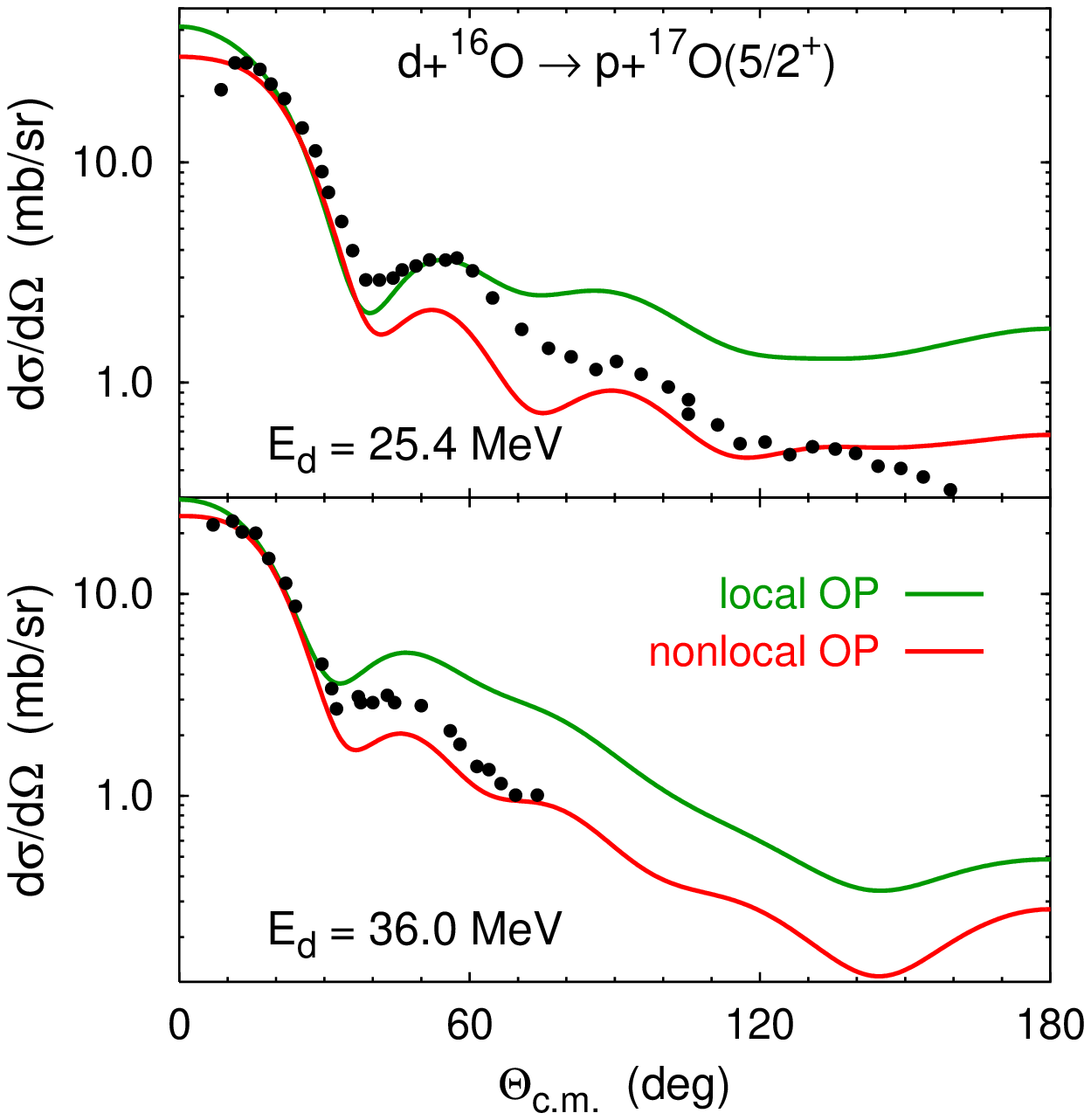} \quad
\includegraphics[width=0.98\columnwidth]{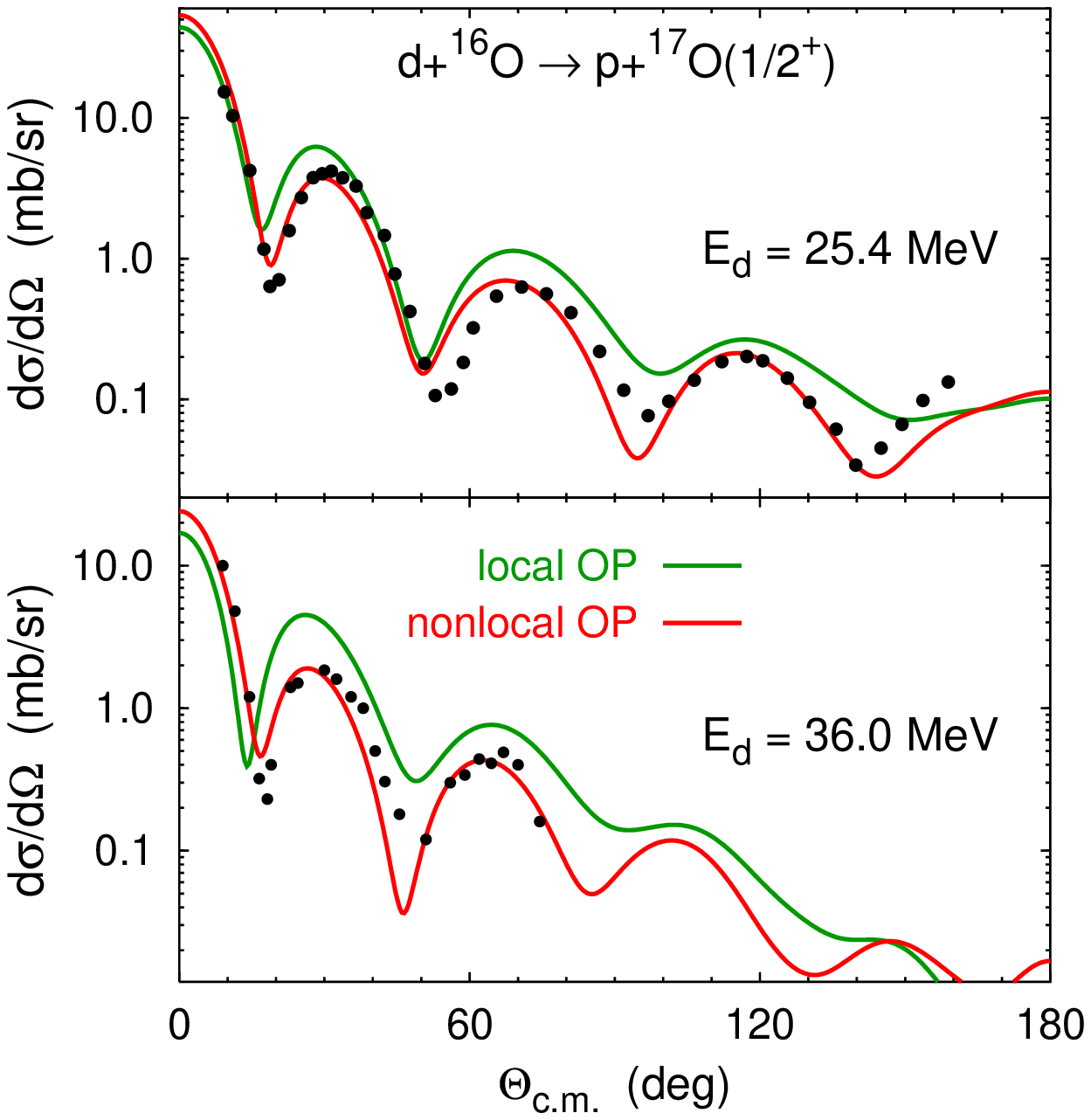}
\caption{\label{fig:O} 
Differential cross section for  
$d + {}^{16}\mathrm{O} \to p+{}^{17}\mathrm{O}$ transfer to
the ${}^{17}\mathrm{O}$ ground state $5/2^+$ (left side)
and excited state  $1/2^+$ (right side)
 at $E_d = 25.4$ and 36.0 MeV calculated with local and nonlocal
optical potentials.
 The experimental data are from Ref.~\cite{dO25-63}.}
\end{figure*}

The screening and renormalization method in the framework of 
momentum-space AGS equations can be applied to the description of 
three-body-like nuclear reactions involving heavier nuclei 
whose interaction with nucleons is described by the optical potentials.
Examples are deuteron  scattering on a stable nucleus
$A$ or proton  scattering on a weakly bound two-body system $(An)$
consisting of the core $A$ and the neutron $n$;  all
elastic, transfer, charge-exchange, and breakup reactions, allowed
by the chosen Hamiltonian, can be calculated on the same footing.
The calculations have been performed for a number of reactions
with the core $A$ ranging from the $\alpha$ particle to  ${}^{58}\mathrm{Ni}$ 
\cite{deltuva:06b,deltuva:09c,cravo:09a}.

In Fig. \ref{fig:ad} we show the differential cross section
for $\alpha$-$d$ breakup reaction at 15~MeV $\alpha$ lab energy 
that is calculated  using three-body $(\alpha,p,n)$ model.
The most important Coulomb effect is the shift of
the  $\alpha p$ $P$-wave resonance position that leads to 
the corresponding changes in the structure of the observables.
The predictions without Coulomb fail completely in accounting for the 
experimental data, while inclusion of the Coulomb moves the peaks of
the differential cross section to the right positions, although the
height of those peaks is not always reproduced,
probably due to deficiencies of the nucleon-$\alpha$ potentials.

The developed technique allowed to test the accuracy of 
traditional approximate nuclear reaction 
approaches like Continuum Discretized Coupled Channels (CDCC) method
\cite{austern:87}
as has been done in Ref. \cite{deltuva:07d} for
 $d+{}^{12}\mathrm{C}$, $d+{}^{58}\mathrm{Ni}$, and  $p+{}^{11}\mathrm{Be}$
reactions. It was found that CDCC is indeed a reliable method to calculate $d+A$
elastic and  breakup cross sections but may lack
accuracy  for transfer reactions such as 
$p+ {}^{11}\mathrm{Be} \to d+ {}^{10}\mathrm{Be}$
and for breakup of one-neutron halo nuclei
$p+ {}^{11}\mathrm{Be} \to p + n + {}^{10}\mathrm{Be}$.
Furthermore, novel dynamical input like energy-dependent \cite{deltuva:09a}
or nonlocal optical potentials  \cite{giannini,giannini2} could be included for 
the first time due to the use of the momentum-space framework.
Especially important  nonlocality effects were found for $(d,p)$ and $(p,d)$ 
transfer reactions involving stable \cite{deltuva:09b} 
as well as  exotic nuclei \cite{deltuva:09d}.

\subsection{Four-nucleon scattering}

Exact description of the four-nucleon scattering is given by the
Faddeev-Yakubovsky equations \cite{yakubovsky:67}
 for the wave-function components
or by the equivalent AGS equations \cite{grassberger:67}.
We use isospin formalism and solve the
symmetrized AGS equations \cite{deltuva:07a}
for the transition operators. Coulomb interaction is included using the
same idea of  screening and renormalization.
Long- and Coulomb-distorted short-range parts
in the scattering amplitudes are  separated \cite{deltuva:07b,deltuva:07c}.
The former is of two-body nature and its $R \to \infty$ limit is known 
analytically. The Coulomb-distorted short-range part is calculated
by solving symmetrized AGS equations numerically where the screened
Coulomb potential is added to the nuclear proton-proton potential;
the $R \to \infty$
limit is reached with sufficient accuracy at finite screening
radii $R$ as demonstrated in Ref.~\cite{deltuva:07b}.

\begin{figure*}[!]
\begin{center}
\includegraphics[width=1.4\columnwidth]{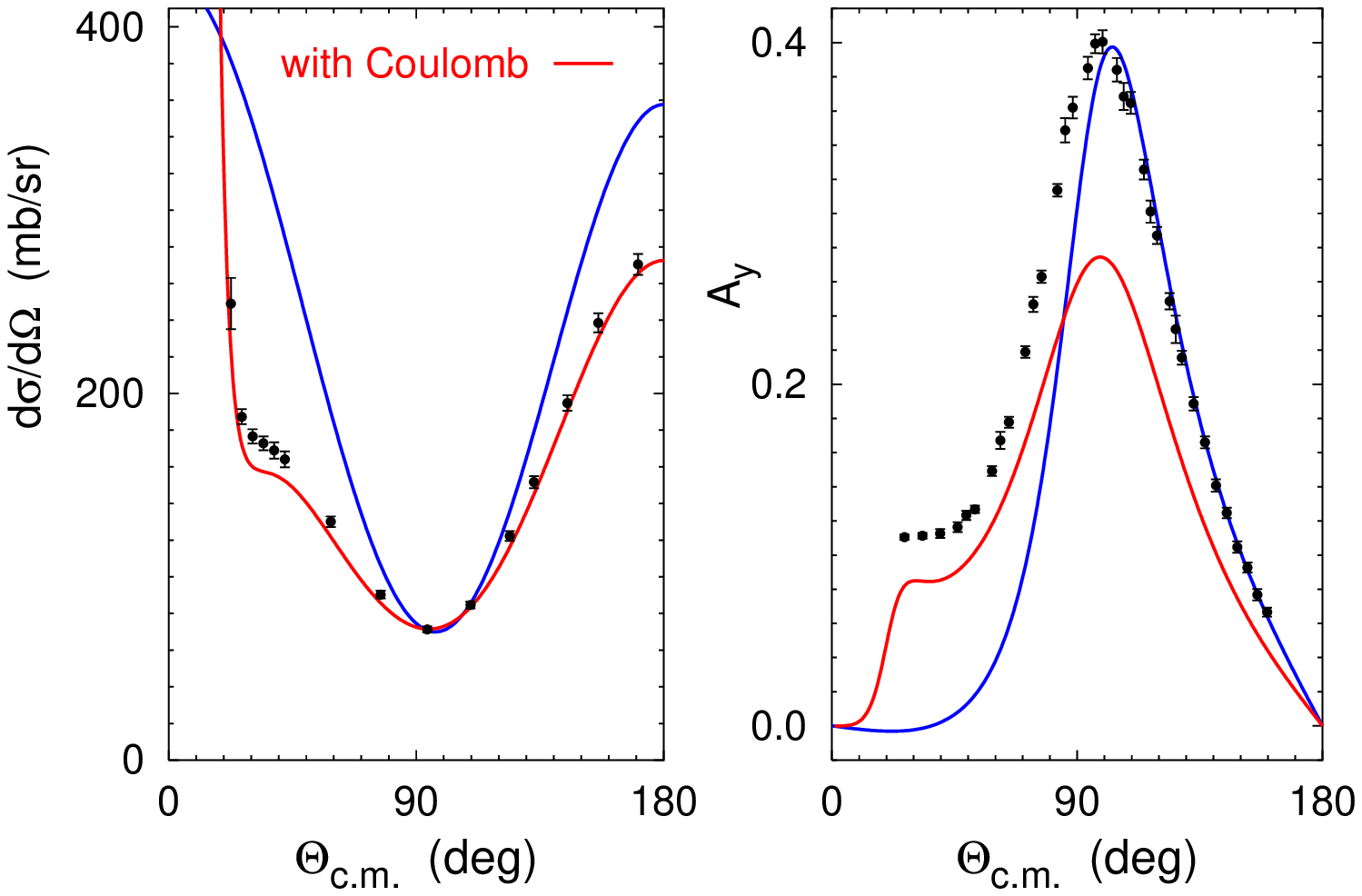}
\end{center}
\caption{\label{fig:pheC}
Differential cross section and proton analyzing power for $p$-${}^3\mathrm{He}$
elastic scattering 
at 4~MeV proton lab energy as functions of the c.m. scattering angle. 
Results including  the Coulomb interaction
(red curves) are compared to results without Coulomb (blue curves).
Hadronic potential is CD Bonn.
The experimental data are from Refs.~\cite{fisher:06,alley:93}. } 
\end{figure*}

The four-nucleon scattering calculations have been performed
so far only below three-body breakup threshold
where the Coulomb effect is extremely important.
This is demonstrated in Fig.~\ref{fig:pheC} for
proton-${}^3\mathrm{He}$ elastic scattering at 4 MeV proton lab energy.
As in $p$-$d$ scattering, the proton  analyzing power is underpredicted
by the theory \cite{deltuva:07b,viviani:01a}.

\begin{figure*}[!]
\begin{center}
\includegraphics[width=1.8\columnwidth]{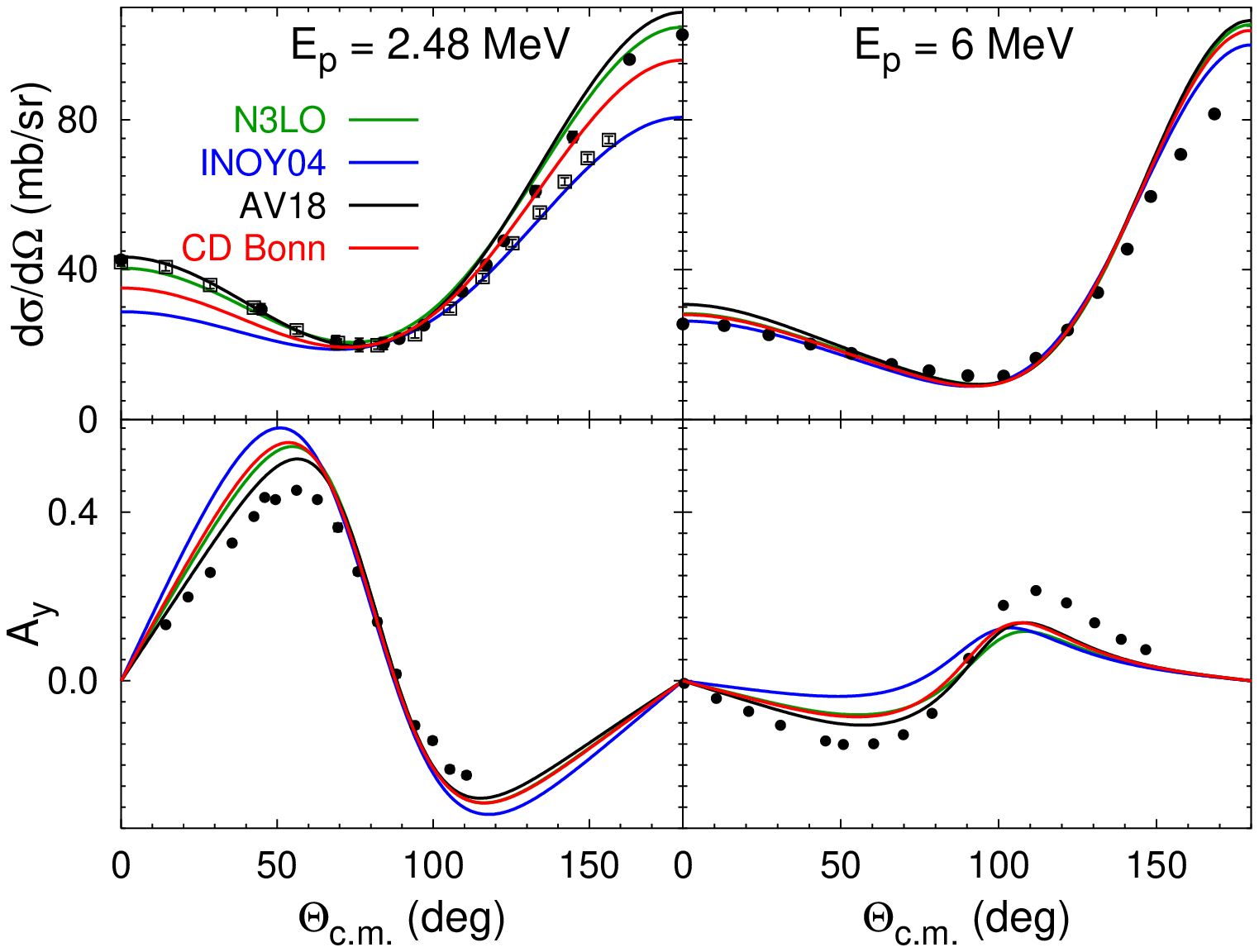}
\end{center}
\caption{\label{fig:pt-nh} 
Differential cross section and proton analyzing power of
$p+{}^3\mathrm{H} \to n + {}^3\mathrm{He}$ reaction at 2.48 and 6 MeV proton lab energy
calculated with various realistic two-nucleon potentials.
The cross section data  are from Refs.~\cite{drosg:78} (circles) and
\cite{jarvis:56} (squares) at 2.48 MeV, and  from Ref.~\cite{wilson:61}
at 6 MeV. $A_y$ data are from Ref.~\cite{doyle:81} at 2.48 MeV
and from Ref.~\cite{jarmer:74} at 6 MeV.}
\end{figure*}

\begin{figure*}[!]
\begin{center}
\includegraphics[width=1.75\columnwidth]{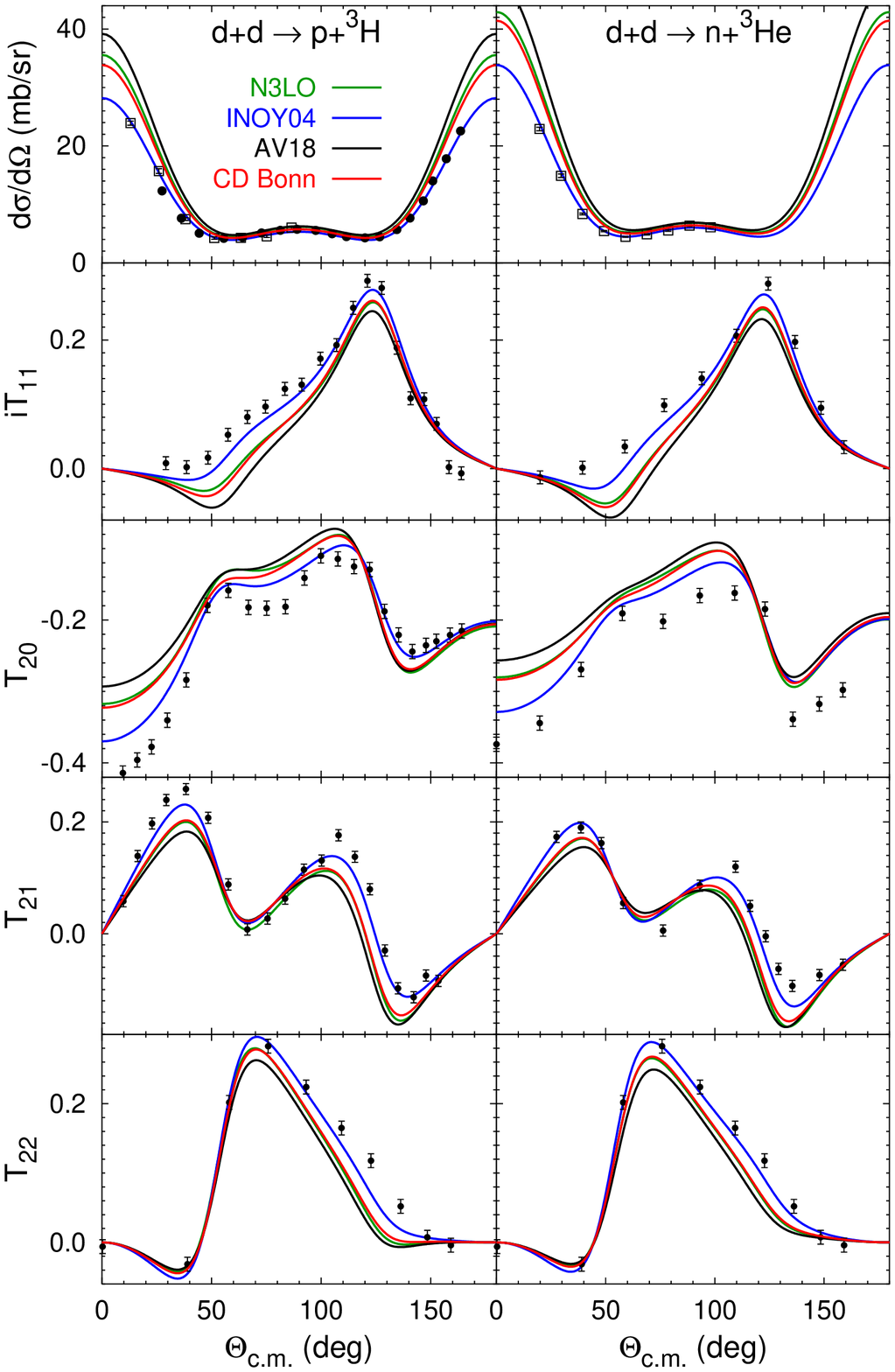}
\end{center}
\caption{\label{fig:dd-ptnh}
Differential cross section and deuteron  analyzing powers of
$d+d \to p+{}^3\mathrm{H}$ and  $d+d \to n+{}^3\mathrm{He}$ reactions at 3 MeV 
deuteron lab energy.
The cross section data  are from Refs.~\cite{blair:48b} (squares)
and \cite{gruebler:72a} (circles) and the analyzing power data are from
Refs.~\cite{gruebler:72a,dries:79a}.}
\end{figure*}

In Figs.~\ref{fig:pt-nh} and \ref{fig:dd-ptnh} we show examples for transfer and
charge-exchange reactions in the four-nucleon system.
The two-nucleon interactions we use are AV18 \cite{wiringa:95a}, 
the one derived from chiral
perturbation theory at next-to-next-to-next-to-leading order
(N3LO) \cite{entem:03a}, CD Bonn \cite{machleidt:01a}, and 
inside nonlocal outside Yukawa (INOY04) potential by
Doleschall \cite{doleschall:04a}.
Although here we do not include a three-nucleon force, its presence is simulated
by using the potential INOY04 that fits both ${}^3\mathrm{He}$ 
 and  ${}^3\mathrm{H}$ experimental
binding energies (7.72 MeV and 8.48 MeV, respectively).
The results for the two-baryon potential CD Bonn + $\Delta$ \cite{deltuva:03c}
allowing for a virtual excitation of a nucleon to a $\Delta$-isobar 
and thereby yielding consistent effective three- and four-nucleon forces
are qualitatively similar and can be found in Ref.~\cite{deltuva:08a}.
Most of the experimental data are quite well described 
\cite{deltuva:07b,deltuva:07c,viviani:01a,lazauskas:09a} at least by some
of the used two-nucleon force models, but  there exist also
several discrepancies, e.g., for the neutron-${}^3\mathrm{H}$ total cross 
section \cite{deltuva:07a,lazauskas:04a} and for the proton analyzing power in the 
proton-${}^3\mathrm{He}$ elastic scattering 
\cite{deltuva:07b,viviani:01a,fisher:06}
and in the  $p+{}^3\mathrm{H} \to n + {}^3\mathrm{He}$ 
charge-exchange reaction \cite{deltuva:07c}.

\section{Summary}

We used the method of screening and renormalization to include
the Coulomb interaction between the charged particles in few-body nuclear reactions.
We demonstrated analytically and numerically that the limit of 
infinite screening radius exists only for renormalized amplitudes.
The short range part of the scattering amplitudes were obtained from
the exact few-body scattering equations that were solved
in the momentum-space framework.
We  obtained fully converged results  for three- and four-nucleon scattering
and for three-body-like nuclear reactions.


\end{document}